\documentstyle[12pt,epsfig]{article}
\newlength{\dinwidth}
\newlength{\dinmargin}
\newcommand{\ba}{\begin{array}}
\newcommand{\ea}{\end{array}}
\newcommand{\beq}{\begin{equation}}
\newcommand{\eeq}{\end{equation}}
\newcommand{\bea}{\begin{eqnarray}}
\newcommand{\eea}{\end{eqnarray}}

\def\S{{\bf S}}

\def\bce{\begin{center}}
\def\ece{\end{center}}

\def\nonu{\nonumber}

\def\pa{\partial}
\def\al{\alpha}
\def\be{\beta}
\def\ga{\gamma}

\def\de{\delta}

\def\la{\lambda}
\def\La{\Lambda}

\def\si{\sigma}

\def\S{{\bf S}}

 \def\unit{\hbox to 3.3pt{\hskip1.3pt \vrule height 7pt width .4pt
\hskip.7pt
\vrule height 7.85pt width .4pt \kern-2.4pt
\hrulefill \kern-3pt
\raise 4pt\hbox{\char'40}}}

\def\eps6{{\displaystyle \mathop{\epsilon}^{6}}{}}

\def\nab6{{\displaystyle \mathop{\nabla}^{6}}{}}

\begin{document}
\thispagestyle{empty}
\addtocounter{page}{-1}
\begin{flushright}
{\tt hep-th/0209128}\\
\end{flushright}
\vspace*{1.3cm}
\centerline{\Large \bf Domain Wall}  
\vskip0.3cm
\centerline{\Large \bf 
from  Gauged  $d=4, {\cal N}=8$ Supergravity:
Part II}
\vspace*{1.5cm}
\centerline{\bf Changhyun Ahn {\rm and} Kyungsung Woo }
\vspace*{1.0cm}
\centerline{\it Department of Physics,
Kyungpook National University, Taegu 702-701 Korea}
\vskip0.3cm
\vspace*{0.8cm}
\centerline{\tt ahn@knu.ac.kr }
\vskip2cm
\centerline{\bf abstract}
\vspace*{0.5cm}

The scalar potentials of the non-semi-simple  
$CSO(p,8-p)$($p=7,6,5$) gaugings of ${\cal N}=8$
supergravity are studied for critical points. The $CSO(7,1)$ gauging
has no $G_2$-invariant critical points, the $CSO(6,2)$ gauging
has  three new $SU(3)$-invariant AdS 
critical points 
and the $CSO(5,3)$ gauging
has no $SO(5)$-invariant critical points.
The scalar potential of $CSO(6,2)$ gauging in four dimensions we discovered 
provides the $SU(3)$ invariant scalar potential
of five dimensional $SO(6)$ gauged supergravity.

The nontrivial effective scalar potential can be written in terms of
the superpotential which can be read off from $A_1$ tensor of the theory.
We discuss first-order domain wall solutions by analyzing the supergravity
scalar-gravity action and using some algebraic relations in a complex
eigenvalue of $A_1$ tensor. We examine domain wall solutions of
$G_2$ sectors of noncompact $SO(7,1)$ and $CSO(7,1)$ gaugings and 
$SU(3)$ sectors of $SO(6,2)$ and $CSO(6,2)$ gaugings. 
They share common features with each sector of
compact $SO(8)$ gauged ${\cal N}=8$ supergravity in four dimensions.

We analyze the scalar potentials of 
the $CSO(p,q,8-p-q)$ gauged supergravity we have found before. 
The $CSO(p,6-p,2)$
gauge theory in four dimensions can be reduced from
the $SO(p,6-p)$ gauge theory in five dimensions. Moreover, 
the $SO(p,5-p)$ gauge theory in seven dimensions reduces to $CSO(p,5-p,3)$
gauge theory in four dimensions.
Similarly, 
$CSO(p,q-p,8-q)$ gauge theories in four dimensions are related to
$SO(p,q-p)$($q=2,3,4,7$) gauge theories in other dimensions.


\baselineskip=18pt
\newpage
\section{Introduction}

The domain wall(DW) and quantum field theory(QFT) correspondence 
is a duality between supergravity compactified on domain wall 
spacetimes(which are locally isometric to Anti-de Sitter(AdS) space 
but different from it globally) and quantum field theories describing
the internal dynamics of branes that live on the boundary of such 
spacetimes. 
Compact gaugings are not the only ones for extended supergravities
but there are  rich structures of non-compact and non-semi-simple gaugings.
These gaugings are crucial in the description of the DW/QFT correspondence
as the compact gauged supergravity has played the role in the AdS/conformal
field theory(CFT) duality(that is a correspondence between a certain gauged 
supergravities and conformal field theories). 

The noncompact and non-semi-simple gauged supergravity theories could be 
obtained in the same way the compact $SO(8)$ gauged supergravity theory.
As a result of the complicated nonlinear tensor structure, one has to prove
that the modified $A_1$ and $A_2$ tensors satisfy a rather complicated
quantities to show the supersymmetry of the theory. A different method 
that uses known results of compact $SO(8)$ gauged supergravity theory
was found to generate noncompact and non-semi-simple gaugings such that
one obtains the full nonlinear structure automatically and both
gauge invariance
and supersymmetry are guaranteed.

In a previous paper, Part I  \cite{aw01} we constructed a superpotential
for known non-compact and non-semi-simple gauged supergravity theories and
by looking at the energy-functional, domain wall solutions were obtained in 
which the role of a superpotential was very important. Moreover, by executing 
two successive $SL(8,{\bf R})$ transformations on the compact gauged 
supergravity theory we described a T-tensor, a superpotential and domain wall
solutions of non-semi-simple $CSO(p,q,8-p-q)$ gaugings.
One considers only scalars which are singlets of subgroup of full isometry
group and is looking 
for critical points of the potential restricted to be a function only
of the singlets. Any critical point of restricted potential is a critical 
point of the original full scalar potential according 
to Schurr's lemma \cite{warner}. In 
Part I, the subgroup was to be $SO(p) \times SO(8-p)$ for $SO(p,8-p)$ and 
$CSO(p,8-p)$ gaugings and 
$SO(p) \times SO(q) \times SO(8-p-q)$ for $CSO(p,q,8-p)$
gaugings. 

There was an attempt \cite{hullwarner}
to study whether any critical points are present in 
$G_2$ sector for $SO(7,1)$ gauging, $SU(3)$ sector for $SO(6,2)$ gauging and
$SO(5)$ sector for $SO(5,3)$ gauging. Only the last one has a critical point
with positive cosmological constant.     
  
In this paper, in section 2, we examine the structure of the  
$G_2$ sector for $SO(7,1)$ gauging, $SU(3)$ sector for $SO(6,2)$ gauging,
$SO(5)$ sector for $SO(5,3)$ gauging and $SO(3) \times SO(3)$ sector for
$SO(4,4)$ gauging.
What we are concentrating on is as follows.
 
$\bullet$ $A_1$ tensor and a superpotential from T-tensor for these gauged
supergravity theories.

In section 3, we repeat the procedure of section 2 for the 
non-semi-simple $CSO(p,8-p)$($p=7,6,5$) gaugings.
 What we are interested in is 

$\bullet$  any critical points of $G_2$ sector for
 $CSO(7,1)$ gauging, $SU(3)$ sector for $CSO(6,2)$ gauging and
$SO(5)$ sector for $CSO(5,3)$ gauging.

$\bullet$  $A_1$ tensor and a superpotential from T-tensor for these 
non-semi-simple $CSO(p,8-p)$ gauged supergravity theories. 

In section 4,  
we obtain domain wall solutions from direct extremization of energy-density 
and in order to arrive this, the observation of the presence of 
some algebraic relations of a superpotential will be crucial  since
without those relations one can not cacel out the unwanted cross terms
in the energy functional. What we describe mainly is as follows.

$\bullet$ Domain wall solutions for non-compact $SO(p,8-p)$($p=7,6$)
and non-semi-simple
$CSO(p,8-p)$ gaugings.

In section 5, 
we analyze the potentials of 
the $CSO(p,q,8-p-q)$ gauged supergravity we have found in \cite{aw01} before. 
The 
$CSO(p,q-p,8-q)$($q=2,3,4,5,6,7$) gauge theories in four dimensions
are related to $SO(p,q-p)$($q=2,3,4,5,6,7$ and $1 \leq p < q$) 
gauge theories
in various higher dimensions.
In section 6, we describe the future directions.
In the appendix, we list the nonzero $A_2$ tensor components in the various 
sectors of given gauged supergravity theories.


\section{The Potentials of $ SO(p, 8-p)$ 
Gauged Supergravity }

We used the $SO(p) \times SO(8-p)$-invariant fourth rank tensor
to generate transformations so that the $SO(p,8-p)$ and $CSO(p,8-p)$ gaugings
are produced in Part I. 
The embedding of $SO(p) \times SO(8-p)$ invariant generator
of $SL(8,{\bf R})$ was such that it corresponds to the $56 \times 56$
$E_7$ generator which is a non-compact $SO(p) \times SO(8-p)$ invariant element
of the $SL(8,{\bf R})$ subalgebra of $E_7$. By introducing the 
projectors onto the corresponding eigenspaces,   
 $SO(p) \times SO(8-p)$-invariant fourth rank tensor can be decomposed into
these projectors. The $\xi$-dependent T-tensor in this case 
\cite{hullplb2,hullcqg,hulljhep}
is described by 
\bea
T_{i}^{\;\;jkl}\left( \xi \right) & =
& t_{i}^{\;\;jkl}-\left( 1-\xi \right)
\left(\overline{u}^{kl}_{\;\;\;\;IJ} + \overline{v}^{klIJ} \right)
\nonumber \\ & & \times \left[ \left( P_{\be}^{IJKL}
+ \frac{1}{2}  P_{\ga}^{IJKL} \right) \left(
u_{im}^{\;\;\;KM}\overline{u}_{\;\;\;LM}^{jm} -
v_{imKM}\overline{v}^{jmLM} \right) \right. \nonumber \\ & &
\left. + P_{\ga}^{IJRS} Z_{RS}^{KLMN} \left(
-v_{imKL}\overline{u}^{jm}_{\;\;\;MN} +
u_{im}^{\;\;\;KL}\overline{v}^{jmMN} \right) \right]
\label{tprime}
\eea
where $ t_{i}^{\;\;jkl}$ in the right hand side
is defined as de Wit-Nicolai T-tensor in compact $SO(8)$ gauging
\bea
t_i^{\;jkl}  = 
\left(\overline{u}^{kl}_{\;\;IJ} +
\overline{v}^{klIJ} \right) \left( u_{im}^{\;\;\;JK} 
\overline{u}^{jm}_{\;\;\;KI}-v_{imJK} \overline{v}^{jmKL} \right)
\nonu
\eea
and we introduce the new quantity $Z_{IJKL}^{MN}$ in terms of
quadratic projectors as follows
\bea
Z_{IJKL}^{MN} = \frac{1}{2} \left[ \left(P_{\al} -P_{\be} \right)_{IJMP}
P_{\ga}^{NPKL} -P_{\ga}^{IJMP} \left(P_{\al} -P_{\be} \right)_{NPKL} \right].
\nonu
\eea
When $\xi=1$, the modified T-tensor reduces to t-tensor in the above.
Projector $P_{\al}(P_{\be})$ projects the $SO(8)$ Lie algebra onto its
$SO(p)(SO(8-p))$ subalgebra while $P_{\ga}$ does onto the remainder 
$SO(8)/(SO(p) \times SO(8-p))$.
Here $\al=-1, \be=p/(8-p)$ and $\ga=(\al+\be)/2$.
The projectors of $SO(p) \times SO(8-p)$-invariant sectors 
are given in the appendix F of Part I \cite{aw01} and corresponding 
$A_1$ and $A_2$ tensors are written as
\bea
A_{1}^{\;\; ij}=-\frac{4}{21}T_{m}^{\;\;ijm},\;\;\;
A_{2l}^{\;\;ijk}=-\frac{4}{3}T_{l}^{\;[ijk]}.
\label{a1a2}
\eea

We describe the potentials of various sectors of $SO(p,8-p)$ and 
$CSO(p,8-p)$
gaugings and are looking for any critical points in the latter. 
In previous paper \cite{aw01}, 
we considered gauged $SO(p,8-p)$
supergravities with $SO(p,8-p)$ gauge symmetry breaking it down to a solution
with symmetry that is some subgroup of $SO(p,8-p)$. That is, 
$SO(p) \times SO(8-p)$ for  $SO(p,8-p)$ gauging.
In this section, we will take the subgroup to be $G_2$ for the $SO(7,1)$
gauging, $SU(3)$ for the $SO(6,2)$ gauging, $SO(5)$ for the $SO(5,3)$ gauging
and $SO(3) \times SO(3)$ for $SO(4,4)$ gauging. 
All these subgroups are  
compact subgroup of noncompact $SO(p,8-p)$. 
Of course, the scalar potentials were obtained already in \cite{hullwarner}
and we will take different approach and see their equivalence. 
The 28-beins for given sectors of gauged supergravity theory in 
(\ref{tprime})
are described completely in terms of some fields \cite{aw}(See the 
appendix). The projectors 
of $SO(p) \times SO(8-p)$ sectors are given in the appendix F of \cite{aw}.
Together with $\xi=-1$, 
the 28-beins for given sectors and the projectors for $SO(p,8-p)$ 
gauged 
supergravity theories, one obtains the modified T-tensor (\ref{tprime}). 
Finally one gets a scalar potential and a superpotential.

\subsection{$G_2$ Sector of $SO(7,1)$ Gauging   }

It is known \cite{warner,dnw,hullwarner} 
that $G_2$-singlet space with a breaking
of the $SO(7)$ gauge subgroup of noncompact $SO(7,1)$ 
into a group which contains $G_2$
may be written as two real parameters $\la$ and $\al$.
The  vacuum expectation value of 56-bein ${\cal V}(x)$
 for the $G_2$-singlet space that is invariant
subspace under a particular $G_2$ subgroup of $SO(7)$ can be 
parametrized by 
\bea
\phi_{ijkl} = \la \cos \al \left( Y^{1\;+}_{ijkl} + Y^{2\;+}_{ijkl} \right)
+ \la \sin \al \left( Y^{1\;-}_{ijkl} + Y^{2\;-}_{ijkl} \right).
\nonu
\eea
Here the completely anti-symmetric self-dual and anti-self-dual
tensors which are invariant under $SO(7)^{+}$ and $SO(7)^{-}$ respectively
are given in terms of $(Y^{1\;+}_{ijkl} + Y^{2\;+}_{ijkl} )$ 
for the
former and
$(Y^{1\;-}_{ijkl} + Y^{2\;-}_{ijkl})$ for the latter
\footnote{Sometimes these tensors are denoted by $C^{+}_{ijkl}$ and 
$C^{-}_{ijkl}$ respectively \cite{dnw}.
Note that $G_2$ is the common subgroup of
$SO(7)^{+}$ and $SO(7)^{-}$. When $\al=0$, it leads to the $SO(7)^{+}$-singlet
space while $\al=\pi/2$ provides  $SO(7)^{-}$-singlet space. 
} where their explicit forms are:
\begin{eqnarray}
  Y^{1\;\pm}_{ijkl} &=& \varepsilon_{\pm} \left[ \; (\de^{1234}_{ijkl} \pm
\de^{5678}_{ijkl})+
 (\de^{1256}_{ijkl} \pm \de^{3478}_{ijkl})+(\de^{3456}_{ijkl}
 \pm \de^{1278}_{ijkl}) \; \right],
\nonu \\
       Y^{2\;\pm}_{ijkl} &=& \varepsilon_{\pm} \left[ -(\de^{1357}_{ijkl}
\pm \de^{2468}_{ijkl})+(\de^{2457}_{ijkl} \pm
\de^{1368}_{ijkl})+(\de^{2367}_{ijkl} \pm \de^{1458}_{ijkl}) +
 (\de^{1467}_{ijkl} \pm \de^{2358}_{ijkl}) \right]
\label{Ys}
\end{eqnarray}
where $\varepsilon_{+}=1$ and $\varepsilon_{-}=i$ and $+$ gives the scalars
and $-$ the pseudo-scalars of ${\cal N}=8$ supergravity. 
The two scalars $\la$ and $\al$ fields in the $G_2$-invariant 
flow parametrize a $G_2$-invariant subspace of the complete
scalar manifold $E_{7(7)}/SU(8)$ in the $d=4, {\cal N}=8$ supergravity.
The 56-bein ${\cal V}(x)$ preserving $G_2$-invariance  
is a $56 \times 56$ matrix
whose elements are some functions of two fields $\la$ and $\al$
by exponentiating the above vacuum expectation value $\phi_{ijkl}$
of $G_2$-singlet space.
Then 28-beins, $u$ and $v$ can be obtained and 
are $28 \times 28$ matrices given in the appendix
A of \cite{aw} together with $\la'=\la$ and $\phi=\al$.

By applying all the data on $u$ and $v$ and the explicit form of
the projectors $P_{\si}^{IJKL}$ of $SO(7)$-invariant sector 
 given in the appendix
F of \cite{aw01} to the equation (\ref{tprime}), 
it turns out that $A_{1}^{\;\;ij}$ tensor for $G_2$ sector of this 
$SO(7,1)$ gauging with the condition $\xi =-1$ has two distinct complex
eigenvalues, $z_1(\la, \al)$ and $z_2(\la, \al)$
with degeneracies 7, 1 respectively and has the following form
\bea
A_{1}^{\;\;ij} & = & \mbox{diag}\left(z_{1},z_{1},z_{1},z_{1},
z_{1},z_{1},z_{1},z_{2}\right)
\nonu
\eea
where the eigenvalues $z_1(\la, \al)$ and $z_2(\la, \al)$ 
are functions of $\la$ and $\al$ as follows
\bea
z_1 & = & \frac{1}{4} e^{-3i\al} \left( e^{i \al} p + q \right)
\left[ 3p^4 q^2 +3 e^{6i \al} p^2 q^4 - 2 e^{i\al} p^3 q \left(
3p^2 +2 q^2 \right)-4 e^{3i\al} p q \left( p^4 -3p^2 q^2 +q^4 \right)  
\right.
 \nonu \\
& & \left. + e^{2i\al} p^2 \left( 3p^4-8p^2 q^2-6q^4  \right)
 -2 e^{5i\al} p q^3 \left( 2p^2 +3 q^2 \right)
 +
e^{4i\al} \left( -6p^4 q^2 -8 p^2 q^4 +3 q^6 \right) \right], \nonu \\
z_{2} & = &
\frac{1}{4} \left( 3 p^7 - 7 e^{-i \al} p^6 q-21 e^{-2 i \al}
p^5 q^2 - 7 e^{-3 i \al} p^4 q^3 - 7 e^{-4 i \al} p^3 q^4 \right. \nonu \\
& & \left. -21
e^{-5 i \al} p^2 q^5 - 7 e^{- 6 i \al} p q^6 + 3 e^{- 7 i \al} q^7 \right)
\label{eigeng2}
\eea
and we denote some hyperbolic functions of $\la$ by the
following quantities which will be used all the times in this paper 
\bea
p \equiv \cosh\left(\frac{\la}{2 \sqrt{2}}\right), \qquad
q \equiv \sinh\left(\frac{\la}{2 \sqrt{2}}\right).
\label{pq}
\eea
The behavior of these eigenvalues of $A_1$ tensor looks similar to
the $G_2$ sector of  compact $SO(8)$ gauging \cite{aw}. 
For $G_2$ sector of
 the non-compact
$SO(7,1)$ gauging, the expressions are more complicated. 
In particular, the magnitude of the eigenvalue $z_2$
plays the role of a superpotential of a scalar potential
which will be discussed in  section 4. 
The scalar potential can be obtained, by putting together all the
components of $A_1$ tensor and $A_2$ tensor written 
explicitly in (\ref{a2}) and (\ref{yso71})
and taking into account the multiplicities, as
\bea
V(\la, \al)  & = &  -g^{2}\left(
\frac{3}{4}|A_{1}^{\;ij}|^{2}-\frac{1}{24}|A_{2i}^{\;\;\;jkl}|^2
\right) \nonu \\
& =& -g^2 \left[
\frac{3}{4} \times \left( 7|z_1|^2 +|z_2|^{2} \right)-\frac{1}{24} \times 6
\left( 7|y_{1,-}|^2 +21|y_{2,-}|^2 +28|y_{3,-}|^2 \right) \right]
\nonu \\
& = &
\frac{1}{2} g^2 \left( c + v s \right)^2 \left[ \left( c + v s \right) \left(
3 c^2 - 8 c v s + 3 v^2 s^2 \right)^2 -14 \left( c - v s \right)
\left( c^2 - 4 c v s + v^2 s^2 \right) \right]
\nonu
\eea
that is exactly the same expression obtained by \cite{hullwarner}
\footnote{The $G_2$ sector of $SO(7,1)$ scalar potential 
can be obtained also by analytic continuation from those sector of 
$SO(8)$ scalar potential
\cite{hullwarner}.
By replacing 56-bein ${\cal V}$ with ${\cal V} E(t)^{-1}$
and scaling by a factor of $e^{2\al t}$,
the potential we are interested in is given by $e^{2\al t} V({\cal V} 
E(t)^{-1})$ at $t=i \pi/(
1+p/(8-p))$. Here $E(t)$ is the $SL(8,{\bf R})$ element and the explicit 
relation between $\xi$ and $t$ is
$\xi=e^
{-(1+\frac{p}{8-p})t}$.
By substituting the transformations 
$
c \rightarrow \frac{1}{\sqrt{2}} \left( c-i s v \right), \;
s v \rightarrow -i \frac{1}{\sqrt{2}} \left( c +i s v \right)
$
with $\al=-1$ and $t=i\pi/8(p=7)$ into the $G_2$ sector of
 $SO(8)$ scalar potential \cite{warner}
$V=2g^2 \left[ (7v^4-7v^2+3) c^3 s^4 +(4v^2-7) v^5 s^7 +c^5 s^2 +
7v^3 c^2 s^5 -3c^3 \right]$ and multiplying the factor $e^{2\al t}=
e^{-i\pi/4}$,
we get the above $G_2$ sector of $SO(7,1)$ scalar potential.}
and we introduce the following
quantities for simplicity
\bea
c \equiv \cosh\left(\frac{\la}{\sqrt{2}}\right), \qquad
s \equiv \sinh\left(\frac{\la}{\sqrt{2}}\right), \qquad
v \equiv \cos \alpha.
\label{csv}
\eea
The analysis in \cite{hullwarner} 
of the $G_2$-invariant critical points of the
$SO(7,1)$ gauging implies  that there is no critical point while the 
$G_2$-invariant compact $SO(8)$ potential possesses four critical points
\cite{warner}:
$SO(8), SO(7)^{+}, SO(7)^{-}$ and $G_2$.

\subsection{$SU(3)$ Sector of $SO(6,2)$ Gauging   }

Similarly the parametrization for the $SU(3)$-singlet
space \cite{warner,hullwarner} 
that has an invariant subspace under a particular
$SU(3)$ subgroup of $SO(6)(=SU(4))$ gauge subgroup of noncompact $SO(6,2)$
can be described by
\bea
\phi_{ijkl} & = &  \la \cos \al  Y^{1\;+}_{ijkl} +
\la \sin \al Y^{1\;-}_{ijkl} + \la' \cos \phi  Y^{2\;+}_{ijkl}
 + \la' \sin \phi Y^{2\;-}_{ijkl}
\nonu
\eea
where the scalar and pseudo-scalar singlets of $SU(3)$ 
are given in (\ref{Ys}) as before.
When we put the constraint of $\la'=\la$ and $\phi=\al$, then 
we get previous $G_2$-invariant sector.
The four scalars $\la, \la', \al$ and $\phi$ fields in the 
$SU(3)$-invariant flow parametrize a $SU(3)$-invariant subspace 
of the complete scalar manifold. 
Then the 56-bein ${\cal V}(x)$ for $SU(3)$-invariance
is a function of $\la, \la', \al$ and $\phi$ and
28-beins $u, v$ are also some functions of these four fields:
we refer to the appendix A of \cite{aw} for explicit relations.
Now we substitute all the expressions of $u$ and $v$ and
the projectors $P_{\si}^{IJKL}$ of $SO(6) \times SO(2)$-invariant sector 
 given in the appendix
F of \cite{aw01} to the defining equation (\ref{tprime}). Then
one obtains that 
 $A_{1}^{\;\;ij}$ tensor for $SU(3)$ sector of this 
$SO(6,2)$ gauging with $\xi =-1$ has
three different complex eigenvalues 
$z_1(\la, \la',\al,\phi), z_2(\la, \la',\al,\phi)$ and 
$z_3(\la, \la', \al, \phi)$
with multiplicities 6, 1, 1
respectively as follows
\bea
A_{1}^{\;\;ij} & = & \mbox{diag}\left(z_{1},z_{1},z_{1},z_{1},
z_{1},z_{1},z_{2},z_{3}\right)
\nonu
\eea
where their explicit dependence on those parameters
are more involved when we compare with the one of  
the $SU(3)$ sector \cite{aw} of  compact $SO(8)$ gauging
but their structure looks similar to those in compact case
and are given
\bea
z_1 & = & \frac{1}{2} e^{-i(\al +2\phi)} \left[
p^2 q r^2 t^2 +e^{4i\phi} p^2 q r^2 t^2 +e^{3i\al} pq^2 r^2 t^2
+ e^{3i\al+ 4i\phi} p q^2 r^2 t^2 \right.
\nonu \\
&& - e^{2i\al} q \left(2p^2 +q^2 \right) r^2 t^2 -
e^{2i(\al+2\phi)} q \left( 2p^2 +q^2 \right) r^2 t^2 -e^{i\al}
p \left( p^2 + 2 q^2 \right) r^2 t^2  \nonu \\
&&  -e^{i(\al +4\phi)} p
\left(p^2 +2q^2 \right) r^2 t^2
-e^{2i\phi} p^2 q \left(r^4 + 4 r^2 t^2 +t^4  \right)
-e^{3i\al +2i\phi} p q^2 \left( r^4 + 4r^2 t^2 +t^4 \right) \nonu \\
&& + e^{i(\al+2\phi)} p \left(-2 q^2
\left( r^4+ t^4 \right) + p^2 \left( r^4 -4 r^2 t^2 +t^4 \right)
\right) \nonu \\
&& \left. + e^{2i(\al+\phi)} \left( -2p^2 q \left( r^4+ t^4 \right) +
q^3 \left( r^4 -4 r^2 t^2 +t^4  \right) \right) \right], \nonu \\
z_2 & = & \frac{1}{2} e^{-3i\al} \left( e^{i\al} p + q \right)
\left( e^{2i\al} p^2 r^4 -4 e^{i\al} p q r^4 +q^2 r^4
- 6 e^{2i(\al +\phi)} p^2
r^2 t^2  \right. \nonu \\
&& \left. -  6 e^{2i\phi} q^2 r^2 t^2 +e^{2i(\al +2\phi)} p^2 t^4 -
4e^{i(\al +4\phi)} p q t^4 + e^{4i\phi} q^2 t^4 \right), \nonu \\
z_3 & = &
\frac{1}{2} e^{-i(3\al +4\phi)} \left( e^{i\al} p +q \right)
\left( e^{2i(\al+2\phi)} p^2 r^4-
4 e^{i(\al+ 4\phi)} p q r^4 +e^{4i\phi} q^2 r^4 -6 e^{2i(\al+\phi)}
p^2 r^2 t^2 \right. \nonu \\
& & \left. -6 e^{2i\phi} q^2 r^2 t^2 +e^{2i\al} p^2 t^4 -
4e^{i\al} p q t^4 +q^2 t^4 \right)
\label{z3info}
\eea
together with the following quantities and (\ref{pq})
\bea
r \equiv \cosh\left(\frac{\la'}{2 \sqrt{2}}\right), \qquad
t \equiv \sinh\left(\frac{\la'}{2 \sqrt{2}}\right).
\label{rt}
\eea
Although the structures of these eigenvalues are more involved,
their degeneracies  resemble the $SU(3)$ sector of
compact $SO(8)$ gauging.  
In this case also, the magnitude of complex $z_3$ will give rise to
a superpotential of a scalar potential which will be discussed later. 
Then the effective nontrivial scalar potential, by plugging the
$A_1$ tensor and $A_2$ tensor given in  (\ref{a2su3}) and 
(\ref{yso62}) 
into the definition of potential and counting the degeneracies correctly,
becomes
\bea
V  & = &  -g^{2}\left(
\frac{3}{4}|A_{1}^{\;ij}|^{2}-\frac{1}{24}|A_{2i}^{\;\;\;jkl}|^2
\right) = -g^2 \left[
\frac{3}{4} \times \left( 6|z_1|^2 +|z_2|^{2}+|z_3|^2 \right) 
 -
\frac{1}{24} \times 6
\left( 3|y_{1,-}|^2  \right. \right. \nonu \\
& & \left. \left. +3|y_{2,-}|^2 +4|y_{3,-}|^2 +12|y_{4,-}|^2+
12|y_{5,-}|^2 +4|y_{6,-}|^2 +6|y_{7,-}|^2 +12|y_{8,-}|^2\right) \right]
\nonu \\
& = &
\frac{1}{2} g^2
\left \{ s'^4
\left[ \left( c +v s \right) \left( 2 x c - \left(x -3
\right) v s \right)^2 -
3 \left( x -1 \right) \left( \left( x+1 \right) c +2 v s
\right)
\right]  \right. \nonu \\
& &  + s'^2 \left[ 2 \left(c + v s \right) \left(
2 c^2 + 2 \left( 3 x-1 \right) c v s - \left( 3 x -
5 \right) v^2 s^2 \right) +
6 \left( \left( x + 1 \right)c -\left( x -3 \right) v s
\right) \right] \nonu \\
& & \left. + 12 v s \right \}
\nonu
\eea
which is the same result of \cite{hullwarner} 
\footnote{
By plugging the transformations of $\la$ and $\al$: 
$
c \rightarrow -i v s, \;
s v \rightarrow  -i c
$
while $\la'$ and $\phi$ 
remain unchanged
with $\al=-1$ and $t=i\pi/4(p=6)$ into the $SU(3)$ sector
of $SO(8)$ scalar potential given in \cite{warner} and multiplying 
the factor $e^{-i\pi/2}$,
the $SU(3)$ sector of  $SO(6,2)$ scalar potential 
can be obtained  by analytic continuation from those sector of
$SO(8)$ scalar potential \cite{hullwarner}.
} and
we introduce the following quantities as well as 
the relations (\ref{csv})
\bea
c' \equiv \cosh\left(\frac{\la'}{\sqrt{2}}\right), \qquad
s' \equiv \sinh\left(\frac{\la'}{\sqrt{2}}\right), \qquad
x \equiv \cos 2 \phi.
\label{c's'v'}
\eea
It was known \cite{hullwarner} 
that there is no $SU(3)$-invariant critical point in $SU(3)$
sector of $SO(6,2)$ gauging. Although 
the compact $SO(8)$ potential has six $SU(3)$-invariant 
critical points \cite{warner}, the
$SO(6,2)$ potential has none. In other words, there are two additional 
critical points, $SU(4)^{-}(=SO(6)^{-})$ 
and $SU(3) \times U(1)$ critical points,
besides the four $G_2$-invariant critical points
we have mentioned in the subsection 2.1.   

\subsection{$SO(5)$ Sector of $SO(5,3)$ Gauging  }

One can construct $SO(5)$-singlets \cite{romans,hullwarner} 
parametrized by
\bea
\phi_{ijkl} & = &  \la \left( X_1^{+} +X_2^{+} + X_3^{+} \right) +
\mu \left( X_1^{+} +X_4^{+} + X_5^{+} \right) +
\rho \left( X_1^{+} - X_6^{+} - X_7^{+} \right)
\label{phiijkl}
\eea
where $\la, \mu$ and $\rho$ characteristic of $SO(5)$-singlets
are three real parameters and
self-dual four-forms are
\bea
& & X_1^{+} = \frac{1}{2} (\de^{1234}_{ijkl} +
\de^{5678}_{ijkl}), \qquad
X_2^{+} =\frac{1}{2}
 (\de^{1256}_{ijkl} + \de^{3478}_{ijkl}), \qquad
X_3^{+}= \frac{1}{2} (\de^{1278}_{ijkl} + \de^{3456}_{ijkl}), \nonu \\
& & X_4^{+} = -\frac{1}{2}(\de^{1357}_{ijkl}
+ \de^{2468}_{ijkl}), \qquad
X_5^{+} = \frac{1}{2} (\de^{1368}_{ijkl} + \de^{2457}_{ijkl}), \qquad
X_6^{+} = \frac{1}{2} (\de^{1458}_{ijkl} + \de^{2367}_{ijkl}),
\nonu \\
& & X_7^{+} = \frac{1}{2}
 (\de^{1467}_{ijkl} + \de^{2358}_{ijkl}).
\label{fourforms}
\eea
In this case, the $SO(5)$ singlet space breaks the $SO(5)$ gauge subgroup
of noncompact $SO(5,3)$ into a group which contains $SO(5)$. 
The three scalars $\la, \mu$ and $\rho$ fields 
in the $SO(5)$-invariant flow parametrize a $SO(5)$-invariant 
subspace of the complete scalar manifold $E_{7(7)}/SU(8)$ in 
$d=4, {\cal N}=8$ supergravity.
The 56-bein preserving $SO(5)$-invariance  
and 28-beins are functions of
three fields $\la, \mu$ and $\rho$ and their explicit form is given in
the appendix B of \cite{aw}.
The eigenvalues of $A_1$ 
tensor are classified by a single real one, $z_1(\la, \mu, \rho)$
which plays the role of a superpotential(which will be studied later)
after we are plugging the expressions of $u$ and $v$ and the projectors
$P_{\si}^{IJKL}$ of $SO(5) \times SO(3)$-invariant sector given in the
appendix F of \cite{aw01} to the equation (\ref{tprime}) with $\xi=-1$:
\bea
A_{1}^{\;\;ij} & = & \mbox{diag}\left(z_{1},z_{1},z_{1},z_{1},
z_{1},z_{1},z_{1},z_{1}\right)
\nonu
\eea
where we write them in terms of
new variables as in the case of  $SO(5)$ sector of compact
$SO(8)$ gauging \cite{romans}
\bea
z_1(\la, \mu, \rho) & = &
\frac{1}{8\sqrt{u v w}} \left( 5 -u^2 v^2 + \mbox{two cyclic
permutations} \right)
\label{superso53}
\eea
where we define 
\bea
u \equiv e^{\la/\sqrt{2}}, \qquad v \equiv e^{\mu/\sqrt{2}},
\qquad w \equiv e^{\rho/\sqrt{2}}.
\label{uvw}
\eea
When we compare with $SO(5)$ sector of $SO(8)$ scalar potential,
there exists a relative sign change in the above.
Finally we will arrive at the scalar potential for $SO(5)$-singlets
by substituting all the components of $A_1$ tensor and 
$A_2$ 
tensor given in  (\ref{a2so5}) and (\ref{yso53})
and taking the multiplicities appropriately:
\bea
V(\la, \mu, \rho) & = & 
  -g^{2}\left(
\frac{3}{4}|A_{1}^{\;ij}|^{2}-\frac{1}{24}|A_{2i}^{\;\;\;jkl}|^2
\right) \nonu \\
&=& 
 -g^2 \left[
\frac{3}{4} \times 8|z_1|^2  
 -
\frac{1}{24} \times 6
\left( 16|y_{1,-}|^2   +16|y_{2,-}|^2 +16|y_{3,-}|^2 +
8|y_{4,-}|^2 \right) \right]
\nonu \\&
=&
\frac{1}{8} g^2 \left( \frac{u^3 v^3}{w} + \frac{10 u v}{w} -2 u v w^3 +
\mbox{two cyclic permutations} - \frac{15}{u v w} \right)
\nonu
\eea
that was observed also in \cite{hullwarner} \footnote{
In this case, we do not need to use Baker-Hausdorff formula because 
the $SO(3)$ action in  the 56-beins ${\cal V}$ commutes with $E(t)^{-1}$ 
for $SO(5,3)$. Therefore the $SO(5,3)$ potential is independent of 
the action of $SO(3)$.
The Lie algebra element generating $E(t)$ can be obtained
by setting $\la=\mu=\rho$. 
By substituting the transformations 
$\frac{\la}{\sqrt{2}} \rightarrow \frac{\la}{\sqrt{2}} -\frac{1}{4} i \pi,
\frac{\mu}{\sqrt{2}} \rightarrow \frac{\mu}{\sqrt{2}} -\frac{1}{4} i \pi,
\frac{\rho}{\sqrt{2}} \rightarrow \frac{\rho}{\sqrt{2}} -\frac{1}{4} i \pi
$ and multiplying the factor 
 $e^{-3\pi i/4}$ into the $SO(5)$ sector of
$SO(8)$ scalar potential \cite{romans} one can get 
this $SO(5)$ sector of $SO(5,3)$ scalar potential \cite{hullwarner}. } 
and note that
the difference from $SO(8)$ potential restricted to $SO(5)$
scalar singlets is the change of sign in the coefficient of $u v/w$
in the above potential.   
It was found that there exists one critical point of this scalar 
potential when $\la=\mu=\rho$(Note that the $SO(5)$-singlet structure
(\ref{phiijkl}) should preserve $SO(5,3)$-invariance characterized by
self-dual antisymmetric four-form tensor $X^{+IJKL}_{5,3}$
written in the appendix A of \cite{aw01} and the condition
$\la=\mu=\rho$ should be satisfied in order to require that
(\ref{phiijkl}) be proportional to  $X^{+IJKL}_{5,3}$) 
and the cosmological constant becomes
$ V=2 \times 3^{1/4} g^2$ with $u=3^{-1/4}$.   
In this subspace the above potential reduces to $SO(5,3)$ scalar potential
$V_{5,3}$ with $\xi=-1$ in 
\cite{aw01} with the identification of $s= -\frac{3}{2\sqrt{2}} \la$ where
$s$ is a scalar field defined in \cite{aw01}.
Note that the $SO(5)$ sector of compact 
$SO(8)$ gauging has two critical points \cite{romans}:
a trivial maximally supersymmetric $SO(8)$ 
critical point and a nonsupersymmetric
$SO(7)$-invariant critical point. All of these are AdS critical points.

\subsection{$SO(3) \times SO(3)$ Sector of $SO(4,4)$ Gauging  }

It is known that $SO(3) \times SO(3)$-singlet space with a breaking
of the $SO(4) \times SO(4)$ into $SO(3) \times SO(3)$ maybe written as
\bea
\phi_{ijkl} = S (\la^{\al} X^{+}_{\al}), \qquad \al = 1, 2, \cdots, 7.
\nonu
\eea
Here the action $S$ is $SO(3) \times SO(3)$ subgroup of $SU(8)$
on its 70-dimensional representation in the space of self-dual four-forms
and is given in \cite{warner1}. Self-dual four forms $X_{\al}^{+}$ are given
in (\ref{fourforms}). The $\la^{\al}$'s that are seven real parameters 
parametrize $SO(3) \times SO(3)$-invariant subspace of 
full scalar manifold in $d=4, {\cal N}=8$ supergravity.
The 56-bein ${\cal V}$ and 28-beins $u,v$ are some functions of these
parameters and they appear in \cite{aw}. 
After we are plugging the expressions of $u$ and $v$ and the projectors
$P_{\si}^{IJKL}$ of $SO(4) \times SO(4)$-invariant sector given in the
appendix F of \cite{aw01} to the equation (\ref{tprime}) with $\xi=-1$,
then one obtains    
$A_1$ 
tensor classified by eight distinct complex ones 
$z_i(i=1, 2, \cdots, 8)$.   
Of course, the structure of these expressions is complicated and the scalar
potential can be obtained as usual. However, it is not very much illuminating 
to present here. It was checked in \cite{hullwarner}
that in this case also there is no critical 
point.


\section{The Potentials of $CSO(p,8-p)$ 
Gauged Supergravity }

In this section, we will take the subgroup to be $G_2$ for the $CSO(7,1)$
gauging, $SU(3)$ for the $CSO(6,2)$ gauging, $SO(5)$ for the 
$CSO(5,3)$ gauging.  
The 28-beins for given sectors of gauged supergravity theory in 
(\ref{tprime})
are described completely in terms of some fields \cite{aw}. The projectors 
of $SO(p) \times SO(8-p)$($p=7,6,5$) 
sectors are given in the appendix F of \cite{aw}.
With $\xi=0$, 28-beins for given sectors and projectors for $CSO(p,8-p)$ 
gauged 
supergravity theory, one obtains the modified T-tensor (\ref{tprime}). 
Finally one gets a new scalar potential by using the definition of scalar
potential given by $A_1$ and $A_2$ tensors.
In particular, the $SU(3)$ sector of $CSO(6,2)$ gauging provides
three AdS critical points which are our new findings.

\subsection{$G_2$ Sector of $CSO(7,1)$ Gauging   }

By applying all the data on $u, v$ which are the same as those in previous 
$SO(7,1)$ gauging and 
the projectors $P_{\si}^{IJKL}$ of $SO(7)$-invariant sector 
given in the appendix
F of \cite{aw01} to (\ref{tprime}), 
$A_{1}$ tensor for $G_2$ sector of this 
$CSO(7,1)$ gauging with the condition $\xi =0$ has two distinct complex
eigenvalues, $z_1(\la, \al)$ and $z_2(\la, \al)$
with degeneracies 7, 1 respectively. 
In this case, $G_2$-singlet space breaks the $SO(7)$ gauge group of
non-semi-simple $CSO(7,1)$ into a group that contains $G_2$.
We emphasize that the only difference between $G_2$ sectors of 
previous $SO(7,1)$ gauging and present $CSO(7,1)$ gauging is that
the parameter  $\xi$ is $-1$ for the former and $0$ for the latter.
Otherwise 28-beins and projectors are the same.
Then the  $A_1$ tensor  has the following form
\bea
A_{1}^{\;\;ij} & = & \mbox{diag}\left(z_{1},z_{1},z_{1},z_{1},
z_{1},z_{1},z_{1},z_{2}\right) \nonu
\eea
where the two distinct eigenvalues $z_1(\la, \al)$ and $z_2(\la, \al)$ are 
given by
\bea
z_1 & = & \frac{1}{8} e^{-3i\al} \left(e^{i\al} p +q \right)
\left( p-e^{i\al} q \right)^2 \nonu \\
& & \times \left[7 p^2 q^2 + 7 e^{4i\al} p^2 q^2 -10 e^{i\al} p q +
10 e^{3i\al} p q + 7 e^{2i\al} \left( p^4 -4 p^2 q^2 +q^4
 \right)  \right], \nonu \\
z_{2} & = &
\frac{7}{8} e^{-7 i \al} \left(-e^{i \al} p+ q \right)^4
\left( e^{i \al} p + q \right)^3
\label{Eigen2}
\eea
where $p$ and $q$ are defined as (\ref{pq}).
The behavior of the eigenvalues of $A_1$ tensor
shares with those sectors in $SO(8)$ and $SO(7,1)$ gaugings.
The superpotential for this theory can be read off from the
expression of $z_2$. 
Now it is straightforward to find out the scalar potential
from $A_1$ tensor and $A_2$ tensor written in (\ref{a2}) and 
(\ref{ycso71})
like we did before
\bea
V(\la, \al)  
& =& -g^2 \left[
\frac{3}{4} \times \left( 7|z_1|^2 +|z_2|^{2} \right)-\frac{1}{24} \times 6
\left( 7|y_{1,0}|^2 +21|y_{2,0}|^2 +28|y_{3,0}|^2 \right) \right]
\nonu \\
& = & \frac{7}{8} \, g^2\, \left( -12 + 7 c^2  - 7  v^2 s^2 \right)
\left( c -v s \right)^3 \left( c + vs \right)^2
\label{potcso71}
\eea
where $c,s$ and $v$ are defined as in (\ref{csv}).
One can obtain also the $G_2$ sector of the $CSO(7,1)$ theory
by analytic continuation as follows: 
As done in obtaining $G_2$ sector of $SO(7,1)$ potential
from those sector of $SO(8)$ scalar potential,
by substituting the transformations \cite{hullwarner} 
\bea
c \rightarrow                    \left( c \cosh 2t - s v \sinh 2t  
\right), \qquad
s v \rightarrow \left( -c \sinh 2t  + s v \cosh 2t   \right)
\nonu
\eea
with $\al=-1$ into the $G_2$ sector of
 $SO(8)$ scalar potential \cite{warner}, 
multiplying the factor $e^{2\al t}$
and taking $t \rightarrow \infty$,
we get the above $G_2$ sector of $CSO(7,1)$ scalar potential.
Note that $\xi=e^
{-(1+\frac{p}{8-p})t}$.
Now we are looking for any critical points of this scalar potential if 
there are.
Differentiating (\ref{potcso71}) with respect to field $\al$, one obtains
\bea
[s(c-s v)^2(c+ s v)(12c-7c^3-(-60+49c^2)s v+ 7c s^2 v^2 +49 s^3 v^3)] 
\sin \al =0.
\nonu
\eea
There exist two possibilities either $\sin \al=0$ or the expression 
in the brackets vanishes.
Let us consider the first case.

$\bullet$ $\sin \al=0$

In terms of $v$, this implies that $v=1$ or $v=-1$. Since $v$ appears
the combination of $v s$ in a scalar potential $V$ (\ref{potcso71}), 
the case
of $v=-1$ maybe obtained from $v=1$ by letting $\la \rightarrow -\la$.
So we need to analyze the case of $v=1$ only. In this subspace, the scalar
potential  (\ref{potcso71}) reduces to 
\bea
V =-\frac{35}{8} g^2 (c-s) = -\frac{35}{8} g^2 e^{-\la/\sqrt{2}}
\nonu
\eea
which does not have any critical points.
Let us describe the second case.

$\bullet$ $\sin \al \neq  0$

Let us change the independent variables in the scalar potential 
$V$ (\ref{potcso71}) as follows:
\bea
A = c, \qquad B = v s
\nonu
\eea
where it is easy to see that this transformation is nonsingular due to 
$\sin \al \neq  0$.
One can find there are no solutions satisfying 
$\pa_A V = \pa_B V =0$ 
where we used the fact that $|A| > |B|$. 

\subsection{$SU(3)$ Sector of $CSO(6,2)$ Gauging   }

In this case, $SU(3)$-singlet space breaks the $SO(6)$ gauge group of
non-semi-simple $CSO(6,2)$ into a group that contains $SU(3)$.
With all the data on $u, v$ and 
the projectors $P_{\si}^{IJKL}$ of $SO(6) \times SO(2)$-invariant sector 
given in the appendix
F of \cite{aw01} that are same as those in 
$SO(6,2)$ gauging, 
$A_{1}^{\;\;ij}$ tensor for $SU(3)$ sector of this 
$CSO(6,2)$ gauging with the condition $\xi =0$ has three distinct complex
eigenvalues
with degeneracies 6, 1, 1 respectively and has the following form
\bea
A_{1}^{\;\;ij} & = & \mbox{diag}\left(z_{1},z_{1},z_{1},z_{1},
z_{1},z_{1},z_{2},z_{3}\right)
\nonu
\eea
where they are given in terms of four paprameters
\bea
z_1 & = & \frac{1}{4} e^{-i(\al +2\phi)} \left( p- e^{i\al} q \right)
\left[ 3 p q r^2 t^2 -3 e^{2i\al} p q r^2 t^2 +3 e^{4i\phi} p q r^2 t^2
 - 3 e^{2i(\al +2\phi)} p q r^2 t^2 \right. \nonu \\
& &  -e^{i\al} r^2 t^2 -
e^{i(\al +4\phi)} r^2 t^2  -e^{2i\phi} p q \left( r^4 +4 r^2 t^2 +
t^4 \right) +e^{2i(\al+\phi)} p q \left( r^4 + 4r^2 t^2+ t^4\right)
\nonu \\
& & \left. + e^{i(\al+2\phi)} \left( 3 r^4 - 4r^2 t^2 +
3 t^4 \right) \right], \nonu  \\
z_2 & = & \frac{3}{4} e^{-3i\al} \left( -e^{i\al} p +q \right)^2
\left( e^{i\al} p + q \right) \left( r^2 -e^{2i\phi} t^2 \right)^2, \nonu \\
z_3 & = &
\frac{3}{4} e^{-i(3\al +4\phi)} \left(-e^{i \al} p +q \right)^2
\left( e^{i \al} p +q\right)
\left( -e^{2i \phi} r^2 + t^2\right)^2
\label{z3info1}
\eea
with (\ref{pq}) and (\ref{rt}).
The scalar potential
from the $A_1$ tensor and $A_2$ tensor given in
(\ref{a2su3}) and (\ref{ycso62}) leads to
\bea
V  & = &   
 -g^2 \left[
\frac{3}{4} \times \left( 6|z_1|^2 +|z_2|^{2}+|z_3|^2 \right) 
 -
\frac{1}{24} \times 6
\left( 3|y_{1,0}|^2  \right. \right. \nonu \\
& & \left. \left. +3|y_{2,0}|^2 +4|y_{3,0}|^2 +12|y_{4,0}|^2+
12|y_{5,0}|^2 +4|y_{6,0}|^2 +6|y_{7,0}|^2 +12|y_{8,0}|^2\right) \right]
\nonu \\
& = & 
\frac{3}{8} \,g^2\, \left( c - s\,v \right) \left[ -2 + s'^{2} \left( x-1
  \right) \right]\left[ 4 + s'^{2}\left( -2 + 3 c^2 - 3 s^2 \, v^2
  \right) \left( x-1 \right)  \right]
\label{potcso62}
\eea
together with (\ref{csv}) and (\ref{c's'v'}). 
By plugging the transformations of $\la$ and $\al$ \cite{hullwarner},
\bea
c \rightarrow                    \left( c \cosh 2t - s v \sinh 2t  
\right), \qquad
s v \rightarrow \left( -c \sinh 2t  + s v \cosh 2t   \right)
\nonu
\eea
with $\al=-1$ into the $SU(3)$ sector
of $SO(8)$ scalar potential \cite{warner}, multiplying 
the factor $e^{-2t}$ and taking the limit of $t \rightarrow \infty$,
the $SU(3)$ sector of  $CSO(6,2)$ scalar potential 
can be obtained  also by analytic continuation from those sector of
$SO(8)$ scalar potential \cite{hullwarner}.
We describe the structure of
critical points of this potential if they exist.
Differentiating (\ref{potcso62}) with respect to field $\al$, one obtains
\bea
&& s\left[ -2 +s'^2(-1+x)\right] \nonu \\
&& \times \left[
4+6 s s'^2(-1+x) \cos \al(c -s \cos \al) +s'^2(-1+x)(
-2 +3c^2-3s^2 \cos^2 \al )
 \right] \sin \al =0.
\nonu
\eea
There exists two possibilities either $\sin \al=0$ or the expression 
in the brackets vanishes.
Let us describe the first case.

\subsubsection{ $\sin \al=0$   }

In terms of $v$, this implies that $v=1$ or $v=-1$. Since $v$ appears
the combination of $v s$ in a scalar potential $V$ (\ref{potcso62}),
the case
of $v=-1$ maybe obtained from $v=1$ by letting $\la \rightarrow -\la$.
So we need to analyze the case of $v=1$ only. In this subspace, the scalar
potential  reduces to 
\bea
V = 
\frac{3}{8} g^2 \left(c-s \right)\left(-2+s'^2(-1+x) \right) 
\left[ 4+s'^2(-1+x)\right] \qquad
\mbox{at} \qquad \al=0.
\label{potcso62reduce}
\eea
Differentiating (\ref{potcso62reduce}) 
with respect to field $\phi$, one obtains
\bea
\frac{\pa V}{\pa \phi} = 
\frac{3}{2} g^2 \left(c-s \right)s'^2 \left(-1+2s'^2 \sin^2 \phi \right) 
\sin 2\phi =0.
\nonu
\eea
There exist four cases we have to consider. Let us describe the case of
$\phi=0, \pi/2$ first.

$\bullet$ $\sin 2\phi=0$

In this subspace , the scalar potential will be
\bea
V & = & -3 g^2 e^{-\la/\sqrt{2}}, 
\;\;\;\;\;\;\;\;\;\;\;\;\;\;\;\;\;\;\;
\;\;\;\;\;\;\;\;\;\;\;\;\;\;\;\;\;\;\;
\mbox{at} \qquad \al=0,\; \phi=0, 
\nonu \\
V & = & \frac{3}{2} g^2 
\left( c-s\right)\left(1+s'^2 \right) 
\left(-2+s'^2 \right), \qquad \mbox{at} \qquad \al=0,\; \phi=\pi/2.
\nonu
\eea
We do not have any critical points in the 
first potential and for the second case
it is easy to see that the conditions of 
$\pa_{\la} V=
\pa_{\la'} V=0$ will provide an imaginary solution for $\la'$
and therefore there are no critical points. 

$\bullet$ $c=s$

There is no real solution for $c=s$ and therefore there is no critical point.

$\bullet$ $s'=0$

The scalar potential becomes
\bea
V=-3 g^2 e^{-\la/\sqrt{2}}
\qquad \mbox{at} \qquad \la' =0
\nonu
\eea
and there is no critical point.

$\bullet$ $-1+2s'^2 \sin^2 \phi=0$

One can substitute $\phi$ or $\la'$ satisfying this condition 
into the potential (\ref{potcso62}) and we get by eliminating $\phi$ 
\bea
V= -\frac{27}{8} g^2 e^{-\la/\sqrt{2}}
\qquad \mbox{at} \qquad \al =0.
\nonu
\eea
We do not have any critical points.
Now we move on the second case.

\subsubsection{ $\sin \al \neq  0$  }

Now let us consider the second case of $\sin \al \neq 0$. Then due to the
negativeness of $ -2 +s'^2(-1+x)$, there are two cases we have to study.
We will describe the first case.

$\bullet$ $s=0$

Let us plug $\la=0$ into the scalar potential (\ref{potcso62}) 
and then we get
\bea
V(\la', \phi) = 
\frac{3}{8} g^2 \left( -2 +s'^2(-1+x)\right) \left( 4+s'^2(-1+x)\right)
\qquad \mbox{at} \qquad \la=0.
\label{newpotential}
\eea 
One can easily get the solutions by differentiating $V$ 
(\ref{newpotential}) with respect to
$\la'$ and $\phi$ and putting zero respectively: 
\bea
i) \; \la'= \pm\frac{1}{\sqrt{2}} \log \left(2+\sqrt{3} \right),\;\; 
\phi =\pm \frac{\pi}{2} \qquad
ii) \; \la'=0, \qquad  iii) \;  \phi=0. 
\nonu
\eea
One can get an extra condition of $\al=\pi/2$(or $3\pi/2$) when we perform 
a differentiation of the scalar potential with respect to $\la$ and
evaluate it at the above critical values. By requiring this should
vanish, one gets $\al=\pi/2$(or $3\pi/2$). 
Now one can evaluate the potential at each critical point.
Now we summarize them as follows \footnote{Before analyzing these analytic 
solutions, there was an attempt to get some of these by a numerical method.
We thank T. Fischbacher to help us.}:
\bea
V & = & -\frac{27}{8} g^2,  \;\;\; \mbox{at} \;\;\; 
 \la=0, \; \la'= \pm\frac{1}{\sqrt{2}} \log \left(2+\sqrt{3} \right), 
\; \al=
\frac{\pi}{2}, \; \phi=\pm \frac{\pi}{2} 
\nonu \\
V & = &  -3 g^2,   \;\;\;\;\;\;  \mbox{at} \;\;\; 
\la=0, \; \la' = \mbox{arbitrary},\; \al=
\frac{\pi}{2}, \; \phi=0, \nonu \\
V  & = & -3 g^2,  \;\;\;\;\;\;  \mbox{at}  \;\;\;
\la=0, \; \la'=0, \; \al = \frac{\pi}{2}, \; \phi=\mbox{arbitrary}. 
\label{criticalcon}
\eea 
In the first critical point in the above, one finds that
the $A_1$ tensor eigenvalues are $7/8$ with six degeneracies and $9/8$ with
two degeneracies and neither eigenvalue satisfies $W=\sqrt{-V/6g^2}$
and so the supersymmetry is completely broken. In the last two critical 
points, 
the $A_1$ tensor eigenvalues are $3/4$ with eight degeneracies that does not 
satisfies $W=\sqrt{-V/6g^2}$ also.
We draw the scalar potential $V(\la',\phi)$ given by (\ref{newpotential})
in Fig. 1 in order to visualize the structure of these critical points.
The four critical points at which the cosmological constants becomes 
$-\frac{27}{8} g^2$ correspond to a local minimum while 
a critical point at which $\phi=0$ has flat direction in $\la'$ direction
and  
a critical point at which $\la'=0$ has flat direction in $\phi$ direction.

At $\la=0$ and $\phi=\frac{\pi}{2}$, the scalar potential
further reduces to 
and is described in Fig. 1
\bea
V(\la') & = & \frac{3}{4} g^2 \cosh^2 
\left( \frac{\la'}{\sqrt{2}} \right) \left( -5 + \cosh 
\left(\sqrt{2} \la' \right)  \right) \nonu \\
&= &
\frac{3}{8} g^2 \left(
p^2 -4p-5  \right), \qquad p \equiv \cosh(4\La), \qquad
\la'=2\sqrt{2} \La
\label{redpot}
\eea
which is proportional to the scalar potential of $SU(3)$ sector of
$SO(6)$ gauging in five dimensions \cite{grw}. In the context of
five dimensional viewpoint, this potential has two AdS critical points.
One is a maximally supersymmetric critical point at $\La=0$ corresponding
to the above critical point at which the potential has $-3g^2 $ in four 
dimensions and 
the other is a nonsupersymmetric critical point at $p=\cosh(4\La)=2$,
corresponding to the critical point at which the potential is
$-\frac{27}{8} g^2$ in four dimensions, 
breaking the $SO(6)$ gauge symmetry into $SU(3) \times U(1)$.
The relevant operators in the four dimensional ${\cal N}=4$ 
super Yang-Mills are mapped to the scalars in the supergravity multiplet.
The existence of an unstable  
nonsupersymmetric $SU(3)$-invariant background of
$AdS_5 \times \S^5$ of type IIB string theory was described in 
\cite{dz,gppz} from the mass spectrum of the low-lying states in this 
$SU(3)$-invariant supergravity solution.    

\begin{figure}
\begin{center}
\begin{minipage}[t]{6.8cm}
\centerline{\hbox{\psfig{file=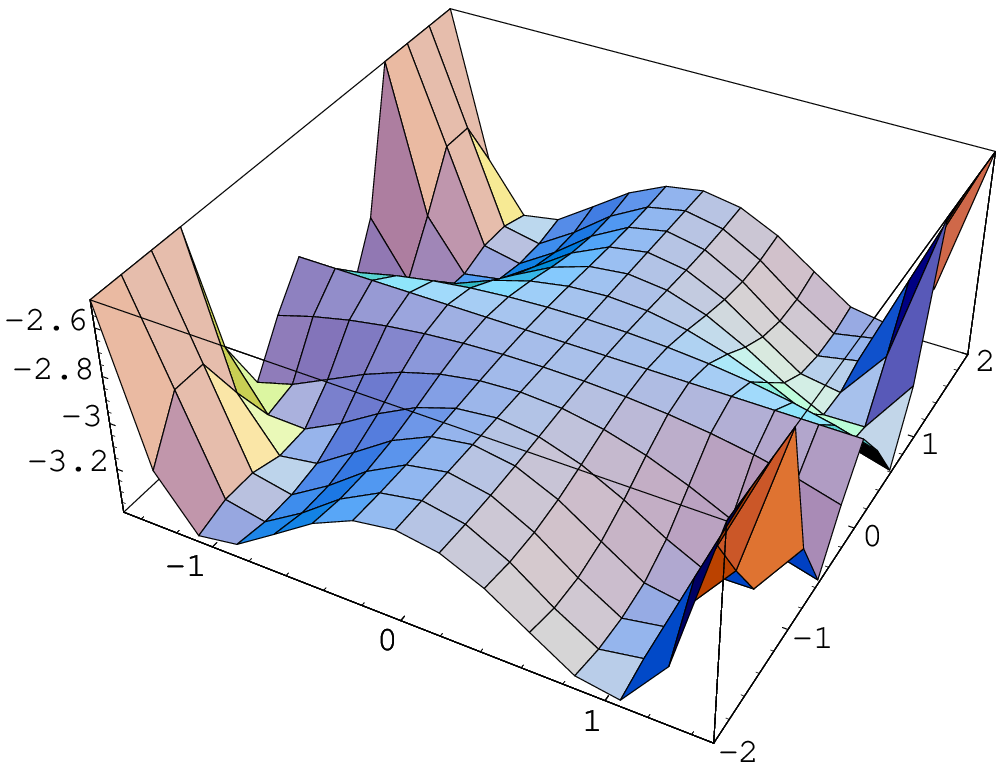,width=6.8cm}}}
\end{minipage}\hspace{1cm}
\begin{minipage}[t]{6.8cm}
\centerline{\hbox{\psfig{file=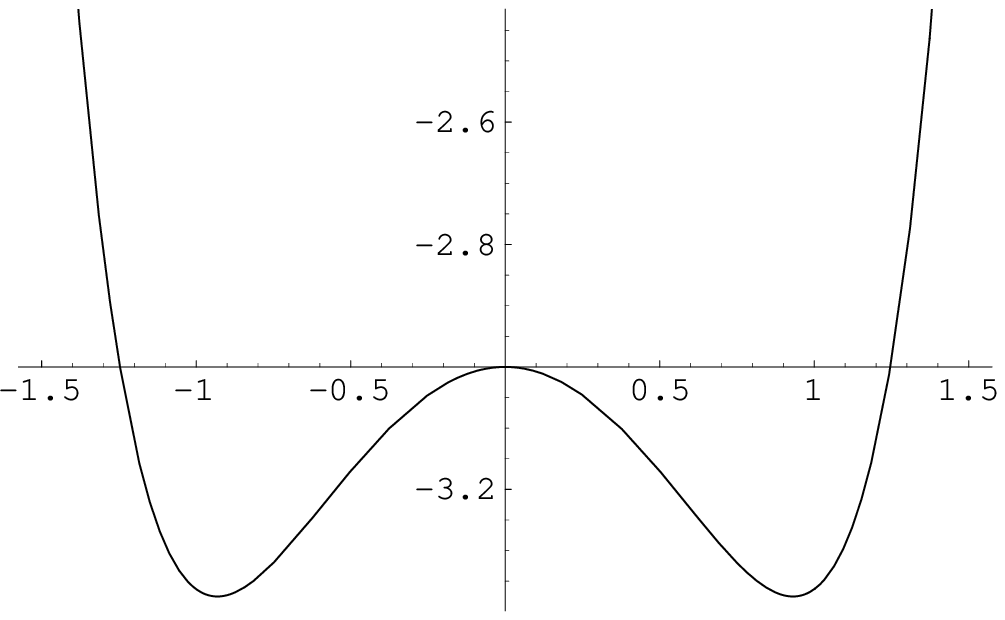,width=6.8cm}}}
\end{minipage}
\caption{\sl The plots of the scalar potential $V(\la',\phi)$(left) at $\la=0, 
\al=\pi/2$  and 
the scalar potential $V(\la')$(right) at $\la=0,\al=\pi/2,\phi=\pi/2$  
in the 4-dimensional gauged supergravity. 
The axes $(\la',\phi)$ in the left are two vevs that 
parametrize the $SU(3)$ invariant manifold in the 28-beins of the theory. 
The four critical points in (\ref{newpotential}) are located at 
$\la'=\pm\frac{1}{\sqrt{2}} \log \left(2+\sqrt{3} \right)\approx \pm 0.93123$
and $\phi=\pm \frac{\pi}{2}\approx \pm 1.5708$
whereas the other critical points where the potentials are flat in these
directions 
are located at an arbitrary point on the line of 
$\phi=0$ or at an arbitrary point on the line of $\la'=0$.
In the right scalar potential (\ref{redpot}) 
we further restricted to the slice of 
$\phi=\pi/2$. 
It turns out that this potential coincides
with the one in $SU(3)$ invariant sector of $SO(6)$ gauging in five 
dimensional supergravity.
We have set the gauge coupling $g$ in the scalar potential as 
$g=1$.  }
\label{fig:spoten}
\end{center}
\end{figure}

Some time ago, a noncompact $SO(6)^{\ast}=SU(3,1)$ gauging in five 
dimensions was constructed \cite{grw1} 
and the scalar potential has a critical point
that breaks the gauge symmetry down to $SU(3) \times U(1)$.
The potential is obtained by replacing $p$ by $-p$ in (\ref{redpot}) and
has a critical point at $\La=0$ at which the potential vanishes.  
Recently, it was shown that dimensionally reducing the $SO(6)^{\ast}$
theory to four dimensional theory and dualizing the graviphoton    
gave the $CSO(6,2)^{\ast}$ gauging which is a non-semi-simple contractions of
$SO(8)^{\ast}=SO(6,2)$ in analogy with the contraction $CSO(p,8-p)$ of
$SO(8)$ \cite{hull02}.  

Let us close this subsection by considering the second case.

$\bullet $ $
4+6 s s'^2(-1+x) \cos \al(c -s \cos \al) +s'^2(-1+x)(
-2 +3c^2-3s^2 \cos^2 \al)=0 $

Let us change the independent variables in the scalar potential
$V$ (\ref{potcso62})  as follows:
\bea
A = c, \qquad B = v s.
\nonu
\eea
Then one can compute the derivatives of $V$ with respect to 
fields $A, B, s'$ and $x$. By requiring that those are vanishing,
one has no real solutions in this case where 
we used that fact that $|A| > |B|$.

\subsection{$SO(5)$ Sector of $CSO(5,3)$ Gauging}

It turns out that the eigenvalues of $A_1$ tensor 
are classified by a single real one, $z_1(\la, \mu, \rho)$ 
\bea
A_{1}^{\;\;ij} & = & \mbox{diag}\left(z_{1},z_{1},z_{1},z_{1},
z_{1},z_{1},z_{1},z_{1}\right), \qquad
z_1  = 
\frac{5}{8 \sqrt{u v w}}
\label{supercso53}
\eea
together with (\ref{uvw}).
The scalar potential is given by with the data of $A_2$ tensor in 
 (\ref{a2so5other}) and (\ref{ycso53}) 
\bea
V(\la, \mu, \rho) & = & 
 -g^2 \left[
\frac{3}{4} \times  8|z_1|^2  
 -
\frac{1}{24} \times 6
\left( 48|y_{1,0}|^2   +8|y_{2,0}|^2  \right) \right]
=
- \frac{15}{8 u v w} g^2.
\nonu
\eea
There is no critical point in this potential.
The $SO(3)$ action in  the 56-beins ${\cal V}$ commutes with $E(t)^{-1}$ 
for $CSO(5,3)$ and the $CSO(5,3)$ potential is independent of 
the action of $SO(3)$.
The Lie algebra element generating $E(t)$ can be obtained
by setting $\la=\mu=\rho$ similarly. 
By substituting the transformations 
\bea
\frac{\la}{\sqrt{2}} \rightarrow \frac{\la}{\sqrt{2}} -\frac{2t}{3},
\qquad
\frac{\mu}{\sqrt{2}} \rightarrow \frac{\mu}{\sqrt{2}} -\frac{2t}{3},
\qquad
\frac{\rho}{\sqrt{2}} \rightarrow \frac{\rho}{\sqrt{2}} -\frac{2t}{3},
\nonu
\eea
 multiplying the factor 
 $e^{-2t}$  and taking $t \rightarrow \infty$ 
into the $SO(5)$ sector of
$SO(8)$ scalar potential \cite{romans} one can get also
this $SO(5)$ sector of $CSO(5,3)$ scalar potential.

\section{ Domain Wall in $SO(p,8-p)$ and $CSO(p,8-p)$
Gaugings }

One of the eigenvalues of $A_1$ tensor for given 
sectors of gauged supergravity theory allows us to write
a superpotential for a scalar potential. In order to find domain-wall
solutions for the theory we have considered so far, it is necessary to
express the energy functional in terms of complete squares in the usual
sense. Since now one can reorganize the scalar potential in terms of
a sum of squares of superpotential and the derivatives of superpotential
with respect to the fields, it leads to the minimization of energy
functional and one gets domain wall solutions without any difficulty. 
This observation is exactly the 
same as the one in the compact $SO(8)$ gauged supergravity theory \cite{aw}.   

\subsection{$G_2$ Sectors of $SO(7,1)$ and $CSO(7,1)$ Gaugings  }

We analyze a particular $G_2$-invariant sector of the 
scalar manifold of gauged ${\cal N}=8$ supergravity. 
The exact information on the supergravity potential
implies  a non-trivial operator algebra in dual field theory.
From the effective scalar potential we have considered so far which
consists of $A_1$ and $A_2$ tensors, one expects that
the superpotential we are looking for maybe encoded in either
$A_1$ or $A_2$ tensor.
One can easily see that one of the eigenvalues of $A_1$ tensor,
that is, $z_2(\la, \al)$
provides a superpotential which is related to
the scalar potential of $SO(7,1)$ or $CSO(7,1)$ gauging as follows:
\bea
V(\la, \al) = \frac{16}{7} g^2 \left| \frac{\pa z_2}{\pa \la} \right|^2 -
6 g^2 |z_2|^2
\label{pot}
\eea
where $z_2$ is given in (\ref{eigeng2}) corresponding to $SO(7,1)$ gauging
 or (\ref{Eigen2}) corresponding to $CSO(7,1)$ gauging. 
This coincides with the one corresponding to $G_2$ sector of compact 
 $SO(8)$ potential \cite{aw,ahnitoh01}. 
The form of this scalar potential in terms of 
a superpotential is quite general for all the cases of $SO(8), SO(7,1)$ and 
$CSO(7,1)$ gaugings.  
Although it seems to
have there is no dependence on the derivative of $z_2$ with respect to
the field $\al$ in the above (\ref{pot}), we have found that
there exists an algebraic relation in complex $z_2$ field
\bea
\pa_{\al} \log |z_2| = 2 \sqrt{2} p q \pa_{\la} \mbox{Arg} z_2
\label{useful}
\eea
implying that one can write the derivative of $z_2$ respect to
$\la$ as two parts. We assume (\ref{pq}) here.
Using this identity, that we have seen in compact $SO(8)$ gauging also,
we will arrive at the
following relation with (\ref{pq})
\bea
V(\la, \al) = 
\frac{16}{7} g^2 \left[ \left(\frac{\pa W}{\pa \la} \right)^2 +
 \frac{1}{8 p^2 q^2} \left( \frac{\pa W}{\pa \al} \right)^2 \right] -
6 g^2 W^2, \qquad W= |z_2|
\nonu
\eea
which is exactly the same as the one in $G_2$ sector of $SO(8)$ potential.
Contrary to the $G_2$ sector of  compact $SO(8)$ potential, one can check that
there are no critical points for these sectors in $SO(7,1)$ and $CSO(7,1)$
potential by differentiating the superpotential $W$ with respect to the 
$\la$ and $\al$ fields.

The Lagrangian of the scalar-gravity sector by adding the scalar potential
we have found to the kinetic terms with vanishing $A_{\mu}^{IJ}$   
can be obtained and is the same as the one in compact $SO(8)$ gauging
except the different potential.
Then 
the resulting Lagrangian has the following form
\bea
\int d^4 x \sqrt{-g} \left ( \frac{1}{2} R - \frac{7}{8}
\pa^{\mu} \la \pa_{\mu} \la - \frac{7}{4} s^2 \pa^{\mu} \al \pa_{\mu} \al - 
V(\la, \al) \right)
\label{action}
\eea
with (\ref{csv}) and $V(\la,\al)$ is a scalar potential for  $SO(7,1)$ gauging
or $CSO(7,1)$ gauging.
To construct domain wall solution corresponding to the supergravity 
description of the nonconformal flow, the metric we are interested in 
is 
\bea
ds^2 = e^{2A(r)} \eta_{\mu \nu} d x^{\mu} dx^{\nu} + e^{2B(r)} dr^2,
\qquad \eta_{\mu \nu} = (-, +, +).
\nonu
\eea
With this ansatz it is straightforward to see
that the equations of motion for the scalar and the metric
from (\ref{action}) are given
\bea
2 \pa^2_r A + 3(\pa_r A)^2 -2 \pa_r A \pa_r B + \frac{7}{8}
(\pa_r \la)^2 +\frac{7}{4} s^2 (\pa_r \al)^2 +e^{2B} V & = & 0, \nonu \\
\pa_r^2 \la +\left(3 \pa_r A-\pa_r B \right) \pa_r \la -\sqrt{2} s c 
(\pa_r \al)^2 -\frac{4}{7} e^{2B} \pa_{\la} V & = & 0, \nonu \\
s^2 \pa_r^2 \al +s^2 \left(3 \pa_r A-\pa_r B \right) \pa_r \al +
\sqrt{2} s c \pa_r \la \pa_r \al -\frac{2}{7} e^{2B} \pa_{\al} V & = & 0. 
\label{secondg2}
\eea  
Then the energy-density per unit area transverse to $r$-direction 
can be obtained. In order to get the first-order differential equations
satisfying the domain-wall, we express the energy-density in terms of
sum of complete sqaures. So one can find out the bound of the energy-density
and it is extremized by the following domain-wall solutions.
Note that in this derivation, we emphasize that the algebraic relation in
(\ref{useful}) was crucial in order to cancel the unwanted terms. 
The flow equations with (\ref{pq}) are \cite{aw,ahnitoh01}
\bea
\partial_{r}\la(r) & = &
\pm \frac{8\sqrt{2}}{7}\,g \, e^{B(r)} \,\partial_{\la} W(\la, \al), \nonu
\\
\partial_{r}\alpha(r) & = &
\pm \frac{\sqrt{2}}{7p^2 q^2}\,g \,e^{B(r)} \, \pa_{\al} W(\la, \al), \nonu \\
\partial_{r} A(r) & = & \mp \sqrt{2} \,g \,e^{B(r)} \, W(\la, \al)
\label{bpsg2}
\eea
where  $W=|z_2|$
and $z_2$ is given in (\ref{eigeng2}) corresponding to $SO(7,1)$ gauging
or (\ref{Eigen2}) corresponding to $CSO(7,1)$ gauging. 
It is straightforward to verify that any solutions of $\la(r), \al(r)$ and
$A(r)$ of (\ref{bpsg2}) satisfy the gravitational and scalar equations
of motion given by the second order equations  (\ref{secondg2}).
We have checked that there are no analytic solutions in (\ref{bpsg2}).

\subsection{$SU(3)$ Sectors of $SO(6,2)$ and $CSO(6,2)$ Gaugings }

We are looking for domain-wall solutions arising in supergravity
theories with nontrivial superpotential defined on the restricted
slice of the scalar manifold.
By similar analysis, one gets the scalar potential and write it in terms of
one of the eigenvalues of $A_1$ tensor
\bea
V(\la, \la', \al, \phi) = 
g^2 \left[ \frac{16}{3}  \left| \frac{\pa z_3}{\pa \la} \right|^2
+ 4  \left| \frac{\pa z_3}{\pa \la'} \right|^2- 6  |z_3|^2 \right].
\nonu
\eea
Here $z_3$ is given in (\ref{z3info}) for $SO(6,2)$ gauging or
(\ref{z3info1}) for $CSO(6,2)$ gauging.
This relation is coincident with the one of $SU(3)$-invariant sector of
$SO(8)$ potential \cite{aw}. In other words, the above structure holds for
$SO(8), SO(6,2)$ and $CSO(6,2)$ gaugings.
At first sight, there are no $\la'$ and $\phi$-derivatives on the
$z_3$. However, one can reexpress those dependences by introducing the
absolute value of $z_3$ as the right superpotential.
It is easy and straightforward 
to check that we have also two algebraic relations
as follows:
\bea
\pa_{\al} \log |z_3| & = & 2 \sqrt{2} p q \pa_{\la} \mbox{Arg} z_3, \nonu \\
\pa_{\phi} \log |z_3| & = & 2 \sqrt{2} r t \pa_{\la'} \mbox{Arg} z_3
\label{useful1}
\eea
providing that the derivative of $z_3$ with respect to $\la$ can be 
decomposed into two parts and the one with respect to $\la'$ into two parts. 
We assume also (\ref{pq}) and (\ref{rt}).
Through these identities one can reexpress the above scalar potential 
as, together with (\ref{pq}) and (\ref{rt}),
\bea
V(\la, \la', \al, \phi) & = &
 g^2 \left[  \frac{16}{3} \left(\partial_{\la}
W \right)^2  + \frac{2}{3p^2 q^2}
\left(\partial_{\al}
W \right)^2 + 4
\left(\partial_{\la'}
W  \right)^2+ \frac{1}{2r^2 t^2} \left(\partial_{\phi}
W \right)^2 - 6  W^2 \right], \nonu \\
W(\la, \la', \al, \phi) & = & |z_3|.
\nonu
\eea
The equations of motion for the scalar and the metric
are given
\bea 
& &
2 \pa_r^2 A +  (3\pa_r A-2\pa_r B) \pa_r A +
\frac{3}{8} (\pa_r \la)^2   +
\frac{3}{4} s^2 (\pa_r \alpha)^2  + 
\frac{1}{2}(\partial_r \la')^2 +  s'^2 ( \pa_r \phi )^2
  + e^{2B} V = 0, \nonu \\ 
& & \pa_r^2 \la +
(3 \pa_r A-\pa_r B) \pa_r \la  - 
 \sqrt{2} s c ( \partial_r \alpha )^2 - \frac{4}{3} e^{2B} \pa_{\la} V   
=0, \nonu \\ 
& &
\pa_r^2 \la' +
(3 \pa_r A -\pa_r B)\pa_r \la'  - 
\sqrt{2} s' c' ( \partial_r \phi )^2 -  e^{2B} \pa_{\la'} V = 0, 
\nonu \\ 
& &
s^2 \pa_r^2 \al + s^2 (3\pa_r A-\pa_r B) \pa_r \al + \sqrt{2} s c \pa_r \al
\pa_r \la -\frac{2}{3} e^{2B} \pa_{\al} V = 0,
\nonu \\ 
&&
s'^2 \pa_r^2 \phi + s'^2 (3\pa_r A -\pa_r B)\pa_r \phi + 
\sqrt{2} s' c' \pa_r \phi
\pa_r \la' -\frac{1}{2} e^{2B} \pa_{\phi} V = 0.
\label{eom1}
\eea 

For given Lagrangian where the kinetic terms are the same as the one in the 
compact $SO(8)$ gauging and given in \cite{aw},
the energy-density can be obtained with the domain wall ansatz we have 
considered. In this case, by using the two relations in (\ref{useful1})
the flow equations \cite{aw} 
with (\ref{pq}) and (\ref{rt}) are 
\bea
\partial_{r}\la(r) & = &
\pm \frac{8\sqrt{2}}{3}g \, e^{B(r)} \, \partial_{\la} 
W(\la,\la', \al,\phi),\nonu
\\
\partial_{r}\la'(r) & = &
\pm 2 \sqrt{2} g \, e^B(r) \, \pa_{\la'} W(\la,\la', \al,\phi), \nonu \\
\partial_{r}\alpha(r) & = & \pm
\frac{\sqrt{2}}{3p^2 q^2} g \, e^{B(r)} 
\, \pa_{\al} W(\la,\la', \al,\phi), \nonu \\
\partial_{r}\phi(r) & = &
\pm \frac{\sqrt{2}}{4r^2 t^2} g \, e^{B(r)} \, \pa_{\phi} W(\la,\la', 
\al,\phi) , \nonu \\
\partial_{r} A(r) & = & \mp \sqrt{2} g \, e^{B(r)} \, W(\la,\la', \al,\phi)
\label{bpssu3}
\eea
where $W=|z_3|$ and we put the $B(r)$ dependence in the right hand side and
$z_3$ is given in (\ref{z3info}) for $SO(6,2)$ gauging or
(\ref{z3info1}) for $CSO(6,2)$ gauging. 
Any solutions of $\la(r), \la'(r), \al(r), \phi(r)$ and
$A(r)$ of (\ref{bpssu3}) satisfy the gravitational and scalar equations
of motion given by the second order equations  (\ref{eom1}).
There are no analytic solutions in (\ref{bpssu3}).


It is easy to see that 
the scalar potential for $SO(5)$ sectors of $SO(5,3)$ and $CSO(5,3)$
gaugings
can be expressed in terms of a superpotential as follows
\bea
V(\la, \mu,\rho) & = & g^2 \left[\frac{32}{5} \left( \partial_{\la}
W \right)^2 + \frac{32}{5}\left( \partial_{\mu} W \right)^2 +
\frac{32}{5}\left( \partial_{\rho} W \right)^2 \right. \nonu \\
& & \left. -\frac{16}{5}
\partial_{\la} W \partial_{\mu} W  -\frac{16}{5}
\partial_{\la} W \partial_{\rho} W
  -\frac{16}{5}   \partial_{\mu} W
\partial_{\rho} W  - 6  W^2 \right],
\label{so5relation}
\eea
where $W=z_1$ is a superpotential (\ref{superso53}) for $SO(5,3)$ gauging
or (\ref{supercso53}) for $CSO(5,3)$ gauging.
The form of this scalar potential in terms of 
a superpotential is quite general for all the cases of $SO(8), SO(5,3)$ and 
$CSO(5,3)$ gaugings. For general $\la, \mu, \rho$, due to the mixed terms in 
the above (\ref{so5relation}) and in the kinetic terms \cite{aw},
there are no domain wall solutions for $SO(5,3)$ gauging 
but under the subspace $\la=\mu=\rho$
we have seen that there is a BPS domain solution \cite{aw01} for 
$SO(5,3)$ gauging.     

%
%
%
%


\section{The Potentials of $CSO(p,q,8-p-q)$ 
Gauged Supergravity }

According to the result of \cite{aw01}, the scalar potential of
$CSO(p,q,8-p-q)$ gauging which is invariant subspace under a particular
$SO(p) \times SO(q) \times SO(8-p-q)$ subgroup of $SO(8)$
can be read off. 
The $CSO(p,q,8-p-q)$ gauging and the $CSO(q,p,8-p-q)$ gauging are
equivalent to each other. So we describe half of them here.

$\bullet $ $CSO(p,6-p ,2)$ gaugings($p=3,4,5$)

Let us consider the scalar potential of $CSO(3,3,2)$ gauging given in terms of
two real scalar fields $\widetilde{m}, \widetilde{n}$ by putting
$\xi=-1$, $\zeta=0$, and $p=3=q$ 
in  the general form of scalar potential of
$CSO(p,q,8-p-q)$ gauging \cite{aw01}.
It is given by 
\bea
V_{3,3,2} =-\frac{3}{4} g^2 e^{-\sqrt{\frac{2}{3}}\widetilde{n}} \left[
\cosh(2 \la) \mp 3 \right], \qquad
\widetilde{m} \rightarrow \sqrt{\frac{3}{2}}  \la
\nonu
\eea
where the $+$ sign in the last term in the above
means the $CSO(6,2)$ gauging because in this case, $\xi=1$ and $\zeta=0$.
At the subspace of $\widetilde{n}=0$, 
the $CSO(3,3,2)$ potential is proportional to  the scalar potential of
noncompact $SO(3,3)$ gauging in five dimensional supergravity \cite{grw}.
This theory in five dimensions
has a de Sitter critical point at which the scalar $\la$
vanishes  and there is no supersymmetry.
Then the  scalar potential which is a $SO(3) \times SO(3)$ invariant sector
of $CSO(3,3,2)$ gauging becomes 
\bea
V=\frac{3}{2} g^2, \qquad \mbox{at} \qquad \la=0.
\label{332cri}
\eea
On the other hand, for $\xi=1, \zeta=0$, 
when the $\la$ vanishes, the scalar potential with $\widetilde{n}=0$
has $V=-3g^2$ we have discussed before
\footnote{The $CSO(6,2)$ scalar potential is $V_{6,2} =
-3 g^2 e^{2s}$ \cite{hullcqg} 
where $s$ is a scalar field that is proportional to
the above $\widetilde{n}$.}.
In subsection 3.2, we have seen the scalar potential in terms of
two fields, $\la'$ and $\phi$. This critical point at which 
$V=-3g^2$ corresponds to 
$SO(3) \times SO(3)$ invariant critical point 
in compact  $SO(6)$
gauged supergravity in five dimensions.
The potential $V_{3,3,2}$ has the exponential roll in the $\widetilde{n}$
direction but is unbounded below in the $\la$ direction \cite{townsend}.
Note that we have found that in \cite{aw01} there exists an analytic
solution for domain wall  of $CSO(3,3,2)$ gauging.

Similarly the scalar potential of $CSO(4,2,2)$ gauging by putting
$\xi=-1$, $\zeta=0$, and $p=4, q=2$ 
in  the general form of scalar potential of
$CSO(p,q,8-p-q)$ gauging is given by
\bea
V_{4,2,2} =- g^2 e^{-\sqrt{\frac{2}{3}}\widetilde{n}} \left(
 e^{2\la} \mp 2 e^{-\la} \right), \qquad
\widetilde{m} \rightarrow \sqrt{3}  \la
\nonu
\eea
where the $+$ sign in the last term in the above
means the $CSO(6,2)$ gauging because in this case, $\xi=1$ and $\zeta=0$.
At the subspace of $\widetilde{n}=0$, 
the $CSO(4,2,2)$ potential
is proportional to  the scalar potential of
$SO(4,2)$ gauging in five dimensional supergravity
\cite{grw} which has no critical points.
The $SO(4) \times SO(2)$ invariant scalars of the $SO(6)$ gauging
in 5-dimensions lead to no new critical points. The scalar potential gives 
$V=-3g^2$ at $\la=0$(in this case, $\xi=1$ and $\zeta=0$).

Finally, 
 the scalar potential of $CSO(5,1,2)$ gauging by putting
$\xi=-1$, $\zeta=0$, and $p=5, q=1$ 
in  the general form of
$CSO(p,q,8-p-q)$ gauging is given by
\bea
V_{5,1,2} = -\frac{1}{8} g^2 e^{-\sqrt{\frac{2}{3}}\widetilde{n}} \left(
15 e^{2\la} \mp 10 e^{-4 \la} - e^{-10 \la}\right), \qquad
\widetilde{m} \rightarrow \sqrt{\frac{15}{2}} \la.
\nonu
\eea
where
 the $+$ sign in the last term 
means the $CSO(6,2)$ gauging because in this case, $\xi=1$ and $\zeta=0$.
At the subspace of $\widetilde{n}=0$, 
the $CSO(5,1,2)$ potential
is proportional to  the scalar potential of
$SO(5,1)$ gauging in five dimensions \cite{grw} which has no critical points.
The $SO(5) $ invariant scalar of the $SO(6)$ gauging
in 5-dimensions($\xi=1$ and $\zeta=0$) leads to 
the  scalar potential with $\widetilde{n}=0$  
\bea
V=-\frac{1}{2} \times 
3^{5/3} g^2 \qquad \mbox{at} \qquad \la=-\frac{1}{6} \log 3.
\label{so5critical}
\eea
Although there are no direct relations  between the supergravity
potentials in four dimensions and in five
dimensions, 
the observation that the $CSO(6,2)$ scalar potential in four dimensions
is related to   the scalar potential for 
$SO(5) $ sector of the $SO(6)$ gauging in  five dimensions 
will provide some hints to understand 
the structure of five dimensional scalar potential in the context of 
full scalar manifold. 
The existence of an unstable  
nonsupersymmetric $SO(5)$-invariant background of
$AdS_5 \times \S^5$ of type IIB string theory was studied in 
\cite{dz,gppz} from the mass spectrum of the low-lying states in this 
$SO(5)$-invariant supergravity solution.   
Moreover, the scalar potential becomes 
$V=-3g^2$ at $\la=0$($\xi=1$ and $\zeta=0$) which is common to
the $CSO(p,6-p ,2)$ gaugings($p=3,4,5$). 
We expect that the $SO(p,6-p)[SO(6)]$ gauge theories in five dimensions
reduce to
the $CSO(p,6-p,2)[CSO(6,2)]$ gauge theories in four dimensions.

$\bullet$ $CSO(p,5-p,3)$ gaugings($p=3,4$)

Let us analyze the scalar potential of $CSO(3,2,3)$ gauging given in terms of
two real scalar fields $\widetilde{m}, \widetilde{n}$ by putting
$\xi=-1$, $\zeta=0$, and $p=3, q=2$ 
in  the general form of scalar potential.
The potential reads 
\bea
V_{3,2,3} = -\frac{3}{8} g^2 e^{-\sqrt{\frac{6}{5}}\widetilde{n}} \left(
 e^{4\la} \mp 4 e^{- \la} \right), \qquad
\widetilde{m} \rightarrow \sqrt{\frac{15}{2}} \la
\nonu
\eea
where one takes $-$ for the $CSO(3,2,3)$ in the second term 
and $+$ for the $CSO(5,3)$ theories.
At the subspace of $\widetilde{n}=0$, 
the $CSO(3,2,3)$ potential is proportional to  the scalar potential of
$SO(3,2)$ gauging in seven dimensional gauged supergravity \cite{ppvw}.
This theory in seven dimensions has no critical point. 
On the other hand, 
the $SO(3) \times SO(2) $ invariant scalar of the $SO(5)$ gauging($\xi=1$ and
$\zeta=0$)
in seven dimensions 
leads to 
the scalar potential  
$V=-\frac{15}{8} g^2$ at $\la=0$ \footnote{
The $CSO(5,3)$ scalar potential \cite{hullcqg} is $V_{5,3} =
-\frac{15}{8} g^2 e^{2s}$ where $s$ is a scalar field that is proportional to
the above $\widetilde{n}$.}. In subsection 3.3, we have seen the scalar 
potential in terms of $\la,\mu$ and $\rho$. At $\la=\mu=\rho=0$,
the scalar potential will coincide with this value.

Similarly the scalar potential of $CSO(4,1,3)$ gauging by putting
$\xi=-1$, $\zeta=0$, and $p=4, q=1$ 
in  the general form of scalar potential is given by
\bea
V_{4,1,3} = -\frac{1}{8} g^2 e^{-\sqrt{\frac{6}{5}}\widetilde{n}} \left(
8 e^{2\la} - e^{-8 \la} \mp  8 e^{-3 \la}\right), \qquad
\widetilde{m} \rightarrow \sqrt{5} \la.
\nonu 
\eea
For $-$ sign in the last term for $CSO(4,1,3)$ gauging
equivalent to $SO(4,1)$ gauged theory in seven dimensions, 
there is no critical point. 
In  seven dimensional gauged supergravity side,
they exist
a local maximum for $\la=0$(at which $V=-\frac{15}{8} g^2$) 
possessing stable and maximally supersymmetric
$SO(5)$ symmetry and 
a local minimum for
$\la=-\frac{1}{5} \log2$ with unstable nonsupersymmetric $SO(4)$ symmetry.
The scalar potential gives 
\bea
V=-\frac{5}{4} \times 2^{3/5} g^2, \qquad
\mbox{at} \qquad \la=-\frac{1}{5} \log2.
\label{413cri}
\eea
Summarizing we expect that the $SO(p,5-p)[SO(5)]$ 
gauge theories in seven dimensions
reduce to
the $CSO(p,5-p,3)[CSO(5,3)]$ gauge theories in four dimensions.

$\bullet$ $CSO(p,4-p, 4)$ gaugings($p=2,3$)

The scalar potential of $CSO(2,2,4)$ gauging by putting
$\xi=-1$, $\zeta=0$, and $p=2=q$ 
in  the general form of scalar potential 
is 
\bea
V_{2,2,4} = \pm g^2 e^{\phi/2}, \qquad 
\widetilde{n} \rightarrow -\frac{\phi}{2\sqrt{2}}.
\nonu
\eea
For $-$ sign in the above it is 
equivalent to $CSO(4,4)$ gauged theory.
At the subspace of $\phi=0$, since the potential has a constant value,
it is a critical point of $SO(2) \times SO(2)$ sector of
$CSO(4,4)$ gauging($\xi=1$ and $\zeta=0$). On the other hand, 
one can interprete the $+$ sign in the above as a noncompact 
$SO(2,2)$ gauged supergravity in seven dimensions that is
a noncompact version of compact $SO(4)$ gauging.
The scalar potential of 
compact $SO(4)$ gauged supergravity was constructed in \cite{ss}. 
By taking the appropriate $SO(2,2)$ metric for T-tensor $T_{ij}$,
it is easy to see that one gets the above potential.  

Similarly the scalar potential of $CSO(3,1,4)$ gauging 
 is given by
\bea
V_{3,1,4} = -\frac{1}{8} g^2 e^{\phi/2} \left(
3 e^{2 \la} \mp 6 e^{-2 \la} -  e^{-6 \la} \right), \qquad
\widetilde{m} \rightarrow \sqrt{3} \la, \qquad \widetilde{n} \rightarrow 
-\frac{\phi}{2\sqrt{2}}
\nonu
\eea
where one takes $-$ for the $CSO(3,1,4)$ in the second term 
and $+$ for the $CSO(4,4)$ theories.
The former is proportional to the  scalar potential of
$SO(3,1)$ gauging in seven dimensions \cite{ss}.
This theory in seven dimensions has no critical point. 
On the other hand, 
the $SO(3) $ invariant scalar of the $SO(4)$ gauging
in seven dimensional supergravity 
leads to 
the scalar potential with $\phi=0$ 
\bea
V=-g^2, \qquad \mbox{at} \qquad  \la=0.
\label{314cri}
\eea
Recall that
the $CSO(4,4)$ scalar potential reads $V_{4,4} =
- g^2 e^{2s}$ where $s$ is a scalar field \cite{hullcqg}.
We expect that the $SO(p,4-p)[SO(4)]$ gauge theories in seven dimensions
reduce to
the $CSO(p,4-p,4)[CSO(4,4)]$ gauge theories in four dimensions.

$\bullet$ $CSO(2,1, 5)$ gauging

Let us consider the scalar potential of $CSO(2,1,5)$ gauging given in terms of
two real scalar fields $\widetilde{m}, \widetilde{n}$ by putting
$\xi=-1$, $\zeta=0$, and $p=2, q=1$ 
in  the general form of scalar potential.
It is given by 
\bea
V_{2,1,5} = \frac{1}{8} g^2 e^{-2\phi} \left(
\pm  e^{-8 \la} +  4 e^{-2 \la}  \right), \qquad
\widetilde{m} \rightarrow  \sqrt{6} \la, \qquad
\widetilde{n} \rightarrow
\frac{\sqrt{6} \phi}{\sqrt{5}}
\nonu
\eea
where the $-$ sign in the first term in the above
means the $CSO(3,5)$ gauging because in this case, $\xi=1$ and $\zeta=0$.
The above $CSO(2,1,5)$ 
scalar potential  is proportional to the  scalar potential of
noncompact $SO(2,1)$ gauging in eight dimensions, that is a noncompact
version of compact $SO(3)$ gauging \cite{ss1,epr},
obtained by taking the appropriate $SO(2,1)$ metric for T-tensor $T_{ij}$.  
This theory in eight dimensions has no critical point. 
On the other hand, 
the $SO(2) $ invariant scalar of the $SO(3)$ gauging
in eight dimensions 
leads to 
the scalar potential(with $\phi=0$) 
$V=- \frac{3}{8} g^2$ at $\la=0$($\xi=1$ and $\zeta=0$).
The $CSO(3,5)$ scalar potential \cite{hullcqg} was $V_{3,5} =
-\frac{3}{8} g^2 e^{2s}$ where $s$ is a scalar field.
We expect that the $SO(2,1)[SO(3)]$ gauge theories in eight dimensions
reduce to
the $CSO(2,1,5)[CSO(3,5)]$ gauge theories in four dimensions.

$\bullet$ $CSO(1, 1, 6)$ gauging

The scalar potential of $CSO(1,1,6)$ gauging by putting
$\xi=-1$, $\zeta=0$, and $p=1=q$ 
in  the general form of scalar potential is  
given by 
\bea
V_{1,1,6} = \frac{1}{8} g^2 e^{\frac{4}{\sqrt{7}} \phi} \left(
  e^{2 \la} +   e^{-2 \la} \pm 2  \right), \qquad
\widetilde{m} \rightarrow \frac{1}{\sqrt{2}} \la, \qquad
\widetilde{n} \rightarrow
-\frac{ 4 \phi}{\sqrt{42}}
\nonu
\eea
where one takes $+$ for the $CSO(1,1,6)$ in the last term 
and $-$ for the $CSO(2,6)$ theories.
By taking the appropriate $SO(1,1)$ metric for T-tensor $T_{ij}$,
the former is proportional to  the scalar potential of
noncompact $SO(1,1)$ gauging in nine dimensions 
which is a noncompact version of
compact $SO(2)$ gauging \cite{cow,nishino,hull02}.
The $SO(1) $ invariant scalar of the $SO(2)$ gauging
in nine dimensions 
leads to one critical point. The scalar potential gives 
$V=0$ at $\la=0$($\xi=1$ and $\zeta=0$) corresponding to  $CSO(2,6)$ 
theory.
The $CSO(2,6)$ scalar potential \cite{hullcqg} was $V_{2,6} = 0$.
Note that the exponential dependence on $\phi$ implies
that $V_{1,1,6}$ can have only critical points at values
$\la=\la_0$ which are critical points of  
$ (e^{2 \la} +   e^{-2 \la} - 2)$(the derivative of this with respect to
$\la$ should vanish at $\la=\la_0$), at which the potential vanishes.
In this case the full potential restricted to
the scalar manifold parametrized by both 
$\phi$ and $\la$ has a critical point  at $\la=0$. 
Note that there exists an analytic domain wall solution 
for this case \cite{aw01}.
The $SO(1,1)[SO(2)]$ gauge theories in nine dimensions
reduce to
the $CSO(1,1,6)[CSO(2,6)]$ gauge theories in four dimensions.

$\bullet $ $CSO(p,7-p ,1)$ gaugings($p=4,5,6$)

Let us consider the scalar potential of $CSO(4,3,1)$ gauging by putting
$\xi=-1$, $\zeta=0$, and $p=4, q=3$ 
in  the general form of scalar potential and
it is given by 
\bea
V_{4,3,1} = -\frac{1}{8} g^2 e^{\frac{2}{7} \phi} \left(
8  e^{2 \la} \mp 24   e^{- \frac{\la}{3}} + 3 e^{-\frac{8 \la}{3}}  \right),
\qquad
\widetilde{m} \rightarrow  \frac{\sqrt{21}}{3} \la, \qquad
\widetilde{n} \rightarrow - \sqrt{\frac{2}{7}} \phi
\nonu
\eea
where the $+$ sign in the last term in the above
means the $CSO(7,1)$ gauging because in this case, $\xi=1$ and $\zeta=0$.
This $CSO(4,3,1)$ potential is proportional to  the scalar potential of
$SO(4,3)$ gauging in four dimensions.
This theory has 
the scalar 
potential(with $\phi=0$) 
is given by, in the $SO(4) \times SO(3)$ invariant sector of 
$CSO(4,3,1)$ gauging, 
\bea
V=\frac{7}{8} \times 2^{8/7} g^2 \qquad 
\mbox{at} \qquad \la=-\frac{3}{7} \log2 .
\label{431cri}
\eea
On the other hand, for $\xi=1, \zeta=0$ there is
a scalar potential with $\phi=0$ 
which has $V=-\frac{35}{8} g^2$ we have discussed 
before(The $CSO(7,1)$ scalar potential \cite{hullcqg} was $V_{7,1} =
-\frac{35}{8} g^2 e^{2s}$ where $s$ is a scalar field).
In subsection 3.1, we have seen the scalar potential $V(\la)$ at the $\al=0$.
At $\la=0$, that potential becomes the same cosmological constant, 
$V=-\frac{35}{8} g^2$. 

Similarly the scalar potential of $CSO(5,2,1)$ gauging 
is given by
\bea
V_{5,2,1} = \frac{5}{8} g^2 e^{\frac{2}{7} \phi} \left(
 -3  e^{2 \la} \mp   4 e^{- \frac{3\la}{2}}   \right),
\qquad
\widetilde{m} \rightarrow  \sqrt{\frac{35}{8}} \la, \qquad
\widetilde{n} \rightarrow -\sqrt{\frac{2}{7}} \phi
\nonu
\eea
where one takes $-$ for the $CSO(5,2,1)$ in the second term 
and $+$ for the $CSO(7,1)$ theories.
The $CSO(5,2,1)$ potential is proportional to  the scalar potential of
$SO(5,2)$ gauging in four dimensions.
This theory has no critical point. 
On the other hand, 
the $SO(5) \times SO(2) $ invariant scalar of the $SO(7)$ gauging
in four dimensions($\xi=1$ and $\zeta=0$) 
leads to 
the scalar potential(with $\phi=0$) 
$V=-\frac{35}{8} g^2$ at $\la=0$.

Finally
the scalar potential of $CSO(6,1,1)$ gauging by putting
$\xi=-1$, $\zeta=0$, and $p=6,q=1$ 
in  the general form of scalar potential is  
given by 
\bea
V_{6,1,1} = \frac{1}{8} g^2 e^{\frac{2}{7} \phi} \left(
 -24  e^{2 \la} \mp   12 e^{-5 \la} +   e^{-12 \la}\right),
\qquad
\widetilde{m} \rightarrow  \sqrt{\frac{21}{2}} \la, \qquad
\widetilde{n} \rightarrow -\sqrt{\frac{2}{7}} \phi
\nonu
\eea
where one takes $-$ for the $CSO(6,1,1)$ in the second term 
and $+$ for the $CSO(7,1)$ theories.
The former is proportional to the  scalar potential of
$SO(6,1)$ gauging in four dimensions.
This theory has no critical point. 
On the other hand, 
the $SO(6) $ invariant scalar of the $SO(7)$ gauging
in four dimensions 
leads to 
the scalar potential  
$V=-\frac{35}{8} g^2$ at $\la=0$($\xi=1$ and $\zeta=0$) 
and for $SO(6)$ invariant sector of $CSO(7,1)$ gauging in four 
dimensions
\bea
V=-7 \times 2^{-4/7}g^2 \qquad \mbox{at} \qquad
\la=-\frac{1}{7} \log 4.
\label{611cri}
\eea
In this case, the $SO(p,7-p)[SO(7)]$ gauge theories in four dimensions
are related to
the $CSO(p,7-p,1)[CSO(7,1)]$ gauge theories in four dimensions.


\section{ Discussions}

In summary,  

$\bullet$ in section 2, we constructed a superpotential from $A_1$ tensor for 
$G_2$ sector for $SO(7,1)$ gauging, $SU(3)$ sector for $SO(6,2)$ gauging,
$SO(5)$ sector for $SO(5,3)$ gauging and $SO(3) \times SO(3)$ sector for
$SO(4,4)$ gauging. In particular, 
the superpotentials are the magnitudes of $z_2$ 
in (\ref{eigeng2}) for $SO(7,1)$ gauging
or (\ref{Eigen2}) for  $CSO(7,1)$ gauging while
they are  given as the magnitudes of $z_3$ 
in (\ref{z3info}) for $SO(6,2)$ gauging or
(\ref{z3info1}) for $CSO(6,2)$ gauging. All these provide the first order 
differential equations. 

$\bullet$ In section 3, we generalized to the 
$G_2$ sector for $CSO(7,1)$ gauging, $SU(3)$ sector for $CSO(6,2)$ gauging,
$SO(5)$ sector for $CSO(5,3)$ gauging. Specially, 
we have discovered three new AdS critical points characterized by
(\ref{criticalcon}) in the 
$SU(3)$ sector for $CSO(6,2)$ gauging, in the four parameter space
of full scalar manifold, preserving the $SU(3)$-invariance.
When we restrict to the subspace parametrized by $\la'$ only,
the scalar potential shows the one in the $SU(3)$-invariant sector
of compact $SO(6)$ gauged supergravity in five-dimensions.   

$\bullet$ In section 4, we obtained 
the first order domain wall solutions for 
$G_2$ sectors (\ref{bpsg2}) for $SO(7,1)$ and $CSO(7,1)$ gaugings and 
$SU(3)$ sectors (\ref{bpssu3}) for $SO(6,2)$ and $CSO(6,2)$ gaugings
by rewriting the scalar potential in terms of a superpotential.
The observation of (\ref{useful}) and (\ref{useful1}) played the role of
elliminating the terms we do not want in the energy-functional.

$\bullet$ In section 5,
we analyzed the behavior of the scalar potentials in 
the $CSO(p,q,8-p-q)$ gauged supergravity theory. 
Along the line of the critical points we have found newly in section 3,
the potential characterized by
(\ref{so5critical}) in the 
$SO(5)$ sector for $CSO(6,2)$ gauging 
was exactly the scalar potential for $SO(5)$-invariant sector of
compact $SO(6)$ gauged supergravity in five-dimensions.   
Also we realized that $CSO(3,3,2)$ gauging 
is given in (\ref{332cri}) and it implies 
the potential for $SO(3) \times SO(3)$-invariant sector of
noncompact $SO(3,3)$ gauged supergravity in five-dimensions.   
There exists a potential
(\ref{413cri})  in the 
$SO(4)$ 
sector for $CSO(5,3)$ gauging in the reduced parameter space corresponding 
to $SO(4)$-invariant sector of
the compact $SO(5)$ gauged supergravity in seven-dimensions.  
We have obtained 
the potential (\ref{431cri}) in the 
$SO(4) \times SO(3)$ 
sector for $CSO(4,3,1)$ gauging
in the reduced parameter space corresponding 
to $SO(4) \times SO(3)$-invariant sector of
the compact $SO(7)$ gauged supergravity in four-dimensions.  
There exists 
(\ref{314cri})  in the 
$SO(3)$ 
sector for $CSO(4,4)$ gauging in the reduced parameter space corresponding 
to $SO(3)$-invariant sector of
the compact $SO(4)$ gauged supergravity in seven-dimensions. 
Also there was 
(\ref{611cri})
in the $SO(6)$  sector for $CSO(7,1)$ gauging corresponding to
the $SO(6)$ invariant sector of the $SO(7)$ gauged supergravity in four 
dimensions.

Let us describe the future directions.  
The scalar potential of gauged ${\cal N}=8$ supergravity
in four dimensions is a function of 70 scalars. We can reduce
the problem by searching for all critical points that reduce 
the gauge/R-symmetry to a group containing a particular
$SO(3)$ subgroup of $SO(8)$. It is known that all of the 
35-dimensional representations of $SO(8)$ contain three 
$SU(3)$-singlets. That is $\bf 8+ 6 +\bar{6}+3 +3 +\bar{3} +
\bar{3} +1 +1 +1$. Under the $SO(3)$ subgroup of $SU(3)$, the irreducible
representation $\bf 6$ of $SU(3)$ breaks into $\bf 5+1$.
Therefore, $SO(3)$-singlet space with a breaking of the $SO(8)$
gauge group into a group which contains $SO(3)$ may be
parametrized by ten real fields.
We expect that there will be new critical points in the $SO(3)$ sector of
guaged ${\cal N}=8$ supergravity
in four dimensions. 
In the context of present work, the 28-beins in those sector can be used
in the $SO(3)$ sector of  $CSO(6,2)$ gauged supergravity.
At least one should find out two AdS critical points. At nonsupersymmetric 
critical point, the potential gives $V=-\frac{3}{2} 
(\frac{25}{2})^{1/3} g^2$
corresponding to $SU(2) \times U(1) \times U(1)$ gauge symmetry in the
five dimensional supergravity and at nonsupersymmetric critical point,
the potential will be $V=-\frac{2^{10/3}}{3} g^2$ corresponding to
$SU(2) \times U(1)$ gauge symmetry in the supergravity side \cite{kpw}.
Note that the critical points in the $SU(2)$ sector of gauged
supergravity in five dimensions are exactly the 
same as the one in the $SO(3)$
sector in that theory \cite{pw}. 
Similarly, it would be interesting to study 
$SU(2)$-singlet space with a breaking of the $SO(8)$
gauge group into a group which contains $SU(2)$.
Among the possible branching rules of ${\bf 3}, \bar{\bf 3},
{\bf 6}, {\bf 8}$ into the
representations of $SU(2)$, the largest singlet structure in $E_{7(7)}$
will provide new critical points. 

When one reduces 11-dimensional supergravity theory to four dimensional
${\cal N}=8$ supergravity, the four dimensional spacetime
is warped by warp factor that provides an understanding of the
different scales of the 11-dimensional solutions. The nonlinear metric ansatz 
in \cite{dnw} provides the explicit formula for the 7-dimensional
inverse metric that is encoded by the warp factor, Killing vectors and
28-beins in four dimensional gauged supergravity theory.
In part I, we identified the 28-beins for $SO(p) \times SO(8-p)$ sectors
with a single vacuum expectation value $\phi$ which depends on the $AdS_4$
radial coordinate $r$. With the insertion of $\xi$-dependence in the
$u,v$, one can easily see that the general expressions for $u,v$ can be 
obtained by simply replacing $\phi$ with $(\phi-t)$ 
because our $u,v$ are related to
${\cal V} E^{-1}(t)$ and we do not need any Baker-Hausdorff formula.
As we have done in the compact gauged supergravity 
\cite{cpw,ahnitoh01,ahnitoh02}, 
one introduces the standard metric of a 7-dimensional ellipsoid 
characterized by the following diagonal matrix 
$
Q_{AB}={\rm diag}\left({\bf 1}_{p}, 
\xi e^{-(1+\be) \phi} {\bf 1}_{8-p}\right),
$   
where $\be=p/(8-p)$.
Then the 7-dimensional metric can be written as $d X^A Q^{-1}_{AB} d X^B$
where the ${\bf R}^8$ coordinate $X^A(A=1, \cdots, 8)$ are constrained on the
unit round 7-sphere, $\sum_A (X^A)^2=1$.
Note that the quadratic form $\Xi^2 = X^A Q_{AB} X^B$ turns to 1 for the 
round 7-sphere with $\phi=0$ and $\xi=1$. The warp factor introduced in 
\cite{gibbonshull,hullwarner88} is nothing but our $\Xi^2$.
Applying the Killing vectors together with the 28-beins $u,v$ to the
metric formula, with the multiplication of $e^{-2t}$, one obtains 
an inverse metric including the warp factor. However, in order to
get the full 7-dimensional metric, one has to separate out the warp factor
from those results.
By plugging the metric with warp factor into the definition of
warp factor, one gets a self-consistent equation for warp factor. 
With this explicit form of warp factor, we will get the final full 
warped 7-dimensional metric corresponding to the one obtained in 
\cite{gibbonshull}.  

With the insertion of $\xi,\zeta$-dependence in the
$u,v$ for $CSO(p,q,8-p-q)$ gauging, 
one can  see that the general expressions for $u,v$ can be 
obtained because our $u,v$ are related to
${\cal V} E^{-1}(t) \times {\cal V} E^{-1}(s) $ 
and we do not need any Baker-Hausdorff formula.
Therefore we replace $\phi$ with $(\phi-t)$ and 
$\chi$ with $(\chi-s)$. 
One introduces the standard metric of a 7-dimensional ellipsoid 
characterized by the following diagonal matrix 
$
Q_{AB}={\rm diag}\left({\bf 1}_{p}, \xi e^{-(1+\be) \phi} {\bf 1}_{q},
\xi \zeta e^{-(1+\be) \phi}  e^{-(1+\be') \chi} {\bf 1}_{8-p-q}\right)$   
where $\be=p/(8-p)$ and $\be'=(p+q)/(8-p-q)$.
Then the 7-dimensional metric can be written as $d X^A Q^{-1}_{AB} d X^B$.
Note that the quadratic form $\Xi^2 = X^A Q_{AB} X^B$ turns to 1 for the 
round 7-sphere with $\phi=0=\chi$ and $\xi=1=\zeta$. 
The warp factor introduced in 
\cite{gibbonshull,hullwarner88} is nothing but our $\Xi^2$.
For $G_2$ sector for $SO(7,1)$ gauging, $SU(3)$ sector for $SO(6,2)$ gauging,
$SO(5)$ sector for $SO(5,3)$ gauging and $SO(3) \times SO(3)$ sector for
$SO(4,4)$ gauging, it would be interesting to develop
the  full warped 7-dimensional metric.

It is natural to ask whether 11-dimensional embedding
of various vacua we have considered of non-compact and non-semi-simple
gauged supergravity can be obtained. In \cite{gibbonshull},
the metric on the 7-dimensional internal space and domain wall in
11-dimensions was found. However, an ansatz for
an 11-dimensional
three-form gauge field is still missing. 
It would be interesting to study the
geometric superpotential,
11-dimensional analog of superpotential we have obtained.
We expect that the nontrivial $r$-dependence of vevs makes
Einstein-Maxwell equations consistent not only at the critical points
but also along the RG flow connecting two critical points.

\section{ Appendix A: Nonzero $A_2$ tensors for given sectors of gauged 
supergravities    }

The nonzero components of $A_2$ tensors can be obtained 
from (\ref{a1a2}) and (\ref{tprime}) by inserting the 28-beins $u,v$
given in the appendix A or B of \cite{aw} and the projectors
in the appendix F of \cite{aw01}. For $SO(p,8-p)$ gauging we put $\xi=-1$
and for $CSO(p,8-p)$ gauging we have $\xi=0$. 
Now we classify them below.

$\bullet$ {\bf $G_2$ sector of $SO(7,1)$ gauging }

In this case, the components of $A_2$ tensor $A_{2,l}^{\;\;\;ijk}$ 
can be represented by
three different fields $y_{i,-}(i=1,2,3)$ with degeneracies 
7,21,28 respectively and given by
\begin{eqnarray}
A_{2,8}^{\;\;\;\;172} & = & A_{2,8}^{\;\;\;\;163} = A_{2,8}^{\;\;\;\;154}
= A_{2,8}^{\;\;\;\;253} = A_{2,8}^{\;\;\;\;246} =
A_{2,8}^{\;\;\;\;374} = A_{2,8}^{\;\;\;\;576} \equiv y_{1,-} \nonumber
\\
A_{2,1}^{\;\;\;\;278} & = & A_{2,1}^{\;\;\;\;368} = A_{2,1}^{\;\;\;\;458} =
A_{2,2}^{\;\;\;\;187} = A_{2,2}^{\;\;\;\;358} = A_{2,2}^{\;\;\;\;486} =
A_{2,3}^{\;\;\;\;186} = A_{2,3}^{\;\;\;\;285} = A_{2,3}^{\;\;\;\;478}
\nonumber \\ & = &
A_{2,4}^{\;\;\;\;185} = A_{2,4}^{\;\;\;\;268} = A_{2,4}^{\;\;\;\;387} =
A_{2,5}^{\;\;\;\;148} = A_{2,5}^{\;\;\;\;238} = A_{2,5}^{\;\;\;\;678} =
A_{2,6}^{\;\;\;\;138} = A_{2,6}^{\;\;\;\;284} \nonumber \\ & = &
A_{2,6}^{\;\;\;\;587} = A_{2,7}^{\;\;\;\;128} = A_{2,7}^{\;\;\;\;348}
= A_{2,7}^{\;\;\;\;568} \equiv y_{2,-} \nonumber \\
A_{2,1}^{\;\;\;\;234} & = & A_{2,1}^{\;\;\;\;256} =
A_{2,1}^{\;\;\;\;375} = A_{2,1}^{\;\;\;\;467} = A_{2,2}^{\;\;\;\;143}
= A_{2,2}^{\;\;\;\;165} = A_{2,2}^{\;\;\;\;367} =
A_{2,2}^{\;\;\;\;457} = A_{2,3}^{\;\;\;\;124} \nonumber \\ & = &
A_{2,3}^{\;\;\;\;157} = A_{2,3}^{\;\;\;\;276} = A_{2,3}^{\;\;\;\;456} =
A_{2,4}^{\;\;\;\;132} = A_{2,4}^{\;\;\;\;176} = A_{2,4}^{\;\;\;\;275}
= A_{2,4}^{\;\;\;\;365} = A_{2,5}^{\;\;\;\;173}  \nonumber \\ & = &
A_{2,5}^{\;\;\;\;126} = A_{2,5}^{\;\;\;\;247} = A_{2,5}^{\;\;\;\;346}
= A_{2,6}^{\;\;\;\;152} = A_{2,6}^{\;\;\;\;147} =
A_{2,6}^{\;\;\;\;237} = A_{2,6}^{\;\;\;\;354} = A_{2,7}^{\;\;\;\;164}
\nonumber \\
& = & A_{2,7}^{\;\;\;\;135} = A_{2,7}^{\;\;\;\;263} =
A_{2,7}^{\;\;\;\;254} \equiv y_{3,-} 
\label{a2}
\end{eqnarray}
where their explicit forms are
\begin{eqnarray}
y_{1,-} & = & \frac{1}{4}\; e^{-i\alpha}\left(p+e^{i\alpha}q\right)^{2} \left[
  -3p^{4}q - 3e^{5i\alpha}pq^{4} + e^{4i\alpha}q^{3} \left( 12p^{2} +
    q^{2} \right) +  e^{i\alpha}p^{3} \left( p^{2} + 12q^{2} \right)
\right. \nonumber \\ & & \left. +
  e^{2i\alpha} \left( 4p^{4}q - 6p^{2}q^{3} \right) + e^{3i\alpha}
  \left( -6p^{3}q^{2} + 4pq^{4} \right) \right], \nonumber \\ 
y_{2,-} & = & -\frac{1}{4}\; e^{-5i\alpha}\left(e^{i\alpha}p+q\right)\left[
  3e^{6i\alpha}p^{4}q^{2} + 4e^{3i\alpha}p^{3}q^{3} + 3p^{2}q^{4} -
  2e^{i\alpha}pq^{3}\left(4p^{2} + q^{2}\right) \right. \nonumber \\ &
& \left. -
  2e^{5i\alpha}p^{3}q\left(p^{2} + 4q^{2}\right) -
  e^{2i\alpha}q^{2}\left(2p^{4} + 8p^{2}q^{2} + q^{4}\right) -
  e^{4i\alpha}p^{2}\left(p^{4} + 8p^{2}q^{2} + 2q^{4}\right)\right],
\nonumber \\
y_{3,-} & = & \frac{1}{4}\; e^{-3i\alpha}\left[-3p^{4}q^{3} -
  3e^{7i\alpha}p^{3}q^{4} + e^{6i\alpha}p^{2}q^{3}\left(4p^{2} +
    3q^{2}\right) + e^{i\alpha}p^{3}q^{2}\left(3p^{2} + 4q^{2}\right)
\right. \nonumber \\ & & \left. + e^{4i\alpha}q^{3}\left(6p^{4} +
    q^{4}\right) + 3e^{5i\alpha}pq^{2}\left(2p^{4} + 4p^{2}q^{2} +
    q^{4}\right) + 3e^{2i\alpha}p^{2}q\left(p^{4} + 4p^{2}q^{2} +
    2q^{4}\right) \right. \nonumber \\ & &\left. +
  e^{3i\alpha}p^{3}\left(p^{4} + 6q^{4} \right) \right]
\label{yso71}
\end{eqnarray}
together with (\ref{pq}). It is clear that $A_{2,l}^{\;\;\;ijk}=-
A_{2,l}^{\;\;\;ikj}$ and  $A_{2,l}^{\;\;\;ijk}=A_{2,l}^{\;\;\;jki}=
A_{2,l}^{\;\;\;kij}$.

$\bullet$ { \bf $G_2$ sector of $CSO(7,1)$ gauging }

With $\xi =0$,  they are classified by
three different fields $y_{i,0}(i=1,2,3)$ with degeneracies 
7,21,28 respectively and given by
(\ref{a2}) with the replacement $y_{i,-} \rightarrow y_{i,0}$.
The redefined expressions are 
\begin{eqnarray}
y_{1,0} &  = & \frac{1}{8} \; e^{-i\alpha}\left(p -
  e^{i\alpha}q\right)^{3} \left(p +
  e^{i\alpha}q \right)^{2}\left(-7pq + 7e^{2i\alpha}pq + e^{i\alpha}
\right), \nonumber \\ 
y_{2,0} & = & -\frac{1}{8}\;e^{-5i\alpha}\left(-e^{i\alpha}p +
  q\right)^{2}\left(e^{i\alpha}p + q\right)\left[7p^{2}q^{2} +
  7e^{4i\alpha}p^{2}q^{2} + 2e^{i\alpha}pq - 2e^{3i\alpha}pq
\right. \nonumber \\ & & \left. -
  e^{2i\alpha}\left(p^{4} + 12p^{2}q^{2} + q^{4}\right)\right],
\nonumber \\
y_{3,0} & = & \frac{1}{8}\;e^{-3i\alpha}\left(p +
  e^{i\alpha}q\right)\left[-7p^{3}q^{3} - 7e^{6i\alpha}p^{3}q^{3} +
  e^{5i\alpha}p^{2}q^{2}\left(11p^{2} + 3q^{2}\right) 
 \right. \nonumber
\\ & & \left. +
  e^{2i\alpha}pq\left(3p^{4} + 9p^{2}q^{2} - 5q^{4}\right) +
  e^{4i\alpha}\left(-5p^{5}q + 9p^{3}q^{3} + 3pq^{5}\right)
\right. \nonumber \\ & & \left. 
+  e^{i\alpha}p^{2}q^{2}\left(3p^{2} + 11q^{2}\right)
+
  e^{3i\alpha}\left(p^{6} -15p^{4}q^{2} -15p^{2}q^{4} + q^{6} \right)
\right] 
\label{ycso71}
\end{eqnarray}
with (\ref{pq}).

$\bullet$ { \bf $SU(3)$ sector of $SO(6,2)$ gauging }

The components of $A_2$ tensor can be represented by
eight different fields $y_{i,-}(i=1,2, \cdots, 8)$ with degeneracies 
3,3,4,12,12,4,6,12 respectively. This looks similar to the compact 
case(that is, same multiplicities and same number of fields)
and they are given by
\begin{eqnarray}
A_{2,7}^{\;\;\;\;128} & = & A_{2,7}^{\;\;\;\;348} =
A_{2,7}^{\;\;\;\;568} \equiv y_{1,-} \nonumber \\
A_{2,8}^{\;\;\;\;172} & = & A_{2,8}^{\;\;\;\;374} =
A_{2,8}^{\;\;\;\;576} \equiv y_{2,-} \nonumber \\
A_{2,7}^{\;\;\;\;164} & = & A_{2,7}^{\;\;\;\;135} = A_{2,7}^{\;\;\;\;263} =
A_{2,7}^{\;\;\;\;254} \equiv y_{3,-} \nonumber \\
A_{2,1}^{\;\;\;\;368} & = & A_{2,1}^{\;\;\;\;458} =
A_{2,2}^{\;\;\;\;358} = A_{2,2}^{\;\;\;\;486} = A_{2,3}^{\;\;\;\;186}
= A_{2,3}^{\;\;\;\;285} = A_{2,4}^{\;\;\;\;185} =
A_{2,4}^{\;\;\;\;268} = A_{2,5}^{\;\;\;\;148} \nonumber \\ & = &
A_{2,5}^{\;\;\;\;238} = A_{2,6}^{\;\;\;\;138} = A_{2,6}^{\;\;\;\;284}
\equiv y_{4,-} \nonumber \\
A_{2,1}^{\;\;\;\;375} & = & A_{2,1}^{\;\;\;\;467} =
A_{2,2}^{\;\;\;\;367} = A_{2,2}^{\;\;\;\;457} = A_{2,3}^{\;\;\;\;157}
= A_{2,3}^{\;\;\;\;276} = A_{2,4}^{\;\;\;\;176} =
A_{2,4}^{\;\;\;\;275} = A_{2,5}^{\;\;\;\;173} \nonumber \\ & = &
A_{2,5}^{\;\;\;\;247} = A_{2,6}^{\;\;\;\;147} = A_{2,6}^{\;\;\;\;237}
\equiv y_{5,-} \nonumber \\
A_{2,8}^{\;\;\;\;163} & = & A_{2,8}^{\;\;\;\;154} = A_{2,8}^{\;\;\;\;253} =
A_{2,8}^{\;\;\;\;246} \equiv y_{6,-} \nonumber \\
A_{2,1}^{\;\;\;\;278} & = & A_{2,2}^{\;\;\;\;187} = A_{2,3}^{\;\;\;\;478} =
A_{2,4}^{\;\;\;\;387} = A_{2,5}^{\;\;\;\;678} = A_{2,6}^{\;\;\;\;587}
\equiv y_{7,-} \nonumber \\
A_{2,1}^{\;\;\;\;234} & = & A_{2,1}^{\;\;\;\;256} = A_{2,2}^{\;\;\;\;143} =
A_{2,2}^{\;\;\;\;165} = A_{2,3}^{\;\;\;\;124} = A_{2,3}^{\;\;\;\;456} =
A_{2,4}^{\;\;\;\;132} = A_{2,4}^{\;\;\;\;365} = A_{2,5}^{\;\;\;\;126}
\nonumber \\ & = & A_{2,5}^{\;\;\;\;346} = A_{2,6}^{\;\;\;\;152} =
A_{2,6}^{\;\;\;\;354} \equiv y_{8,-} 
\label{a2su3}
\end{eqnarray}
where eight fields are 
\begin{eqnarray}
y_{1,-} & = & -\frac{1}{2}\;e^{-i\left(\alpha + 4\phi
  \right)}\left[e^{4i\phi}p^{2}qr^{4} + e^{i\left(3\alpha + 4\phi
    \right)}pq^{2}r^{4} - e^{2i\left(\alpha + 2\phi
    \right)}q\left(2p^{2} + q^{2}\right)r^{4} \right. \nonumber \\ & &
\left. - e^{i\left(\alpha +
      4\phi \right)}p\left(p^{2} + 2q^{2}\right)r^{4} -
  6e^{2i\phi}p^{2}qr^{2}t^{2} - 6e^{i\left(3\alpha + 2\phi
    \right)}pq^{2}r^{2}t^{2} \right. \nonumber \\ & & \left. -
  2e^{2i\left(\alpha + \phi \right)}q\left(2p^{2} +
    q^{2}\right)r^{2}t^{2} - 2e^{i(\alpha + 2\phi)}p\left(p^{2} +
    2q^{2}\right)r^{2}t^{2} + p^{2}qt^{4} + e^{3i\alpha}pq^{2}t^{4}
\right. \nonumber \\ & & \left. - e^{2i\alpha}q\left(2p^{2} +
    q^{2}\right)t^{4} - e^{i\alpha}p\left(p^{2} +
    2q^{2}\right)t^{4}\right],  \nonumber \\ 
y_{2,-} & = & -\frac{1}{2}\;e^{-i\alpha}\left[p^{2}qr^{4} +
  e^{3i\alpha}pq^{2}r^{4} - e^{2i\alpha}q\left(2p^{2} +
    q^{2}\right)r^{4} - e^{i\alpha}p\left(p^{2} + 2q^{2}\right)r^{4}
\right. \nonumber \\ & & \left. -
  6e^{2i\phi}p^{2}qr^{2}t^{2} - 6e^{i\left(3\alpha + 2\phi
    \right)}pq^{2}r^{2}t^{2} - 2e^{2i\left(\alpha + \phi
    \right)}q\left(2p^{2} + q^{2}\right)r^{2}t^{2} \right. \nonumber
\\ & & \left. - 2e^{i\left(\alpha
    + 2\phi \right)}p\left(p^{2} + 2q^{2}\right)r^{2}t^{2} +
e^{4i\phi}p^{2}qt^{4} + e^{i\left(3\alpha + 4\phi \right)}pq^{2}t^{4}
- e^{2i\left(\alpha + 2\phi \right)}q\left(2p^{2} + q^{2}\right)t^{4}
\right. \nonumber \\ & & \left. - e^{i\left(\alpha + 4\phi
    \right)}p\left(p^{2} + 2q^{2}\right)t^{4}\right], \nonumber \\ 
y_{3,-} & = & -\frac{1}{2}\;e^{-3i\alpha}\left(p +
  e^{i\alpha}q\right)rt\left[e^{4i\phi}p^{2}r^{2} - 4e^{i \left(\alpha +
    4\phi \right)}pqr^{2} + e^{2i\left(\alpha + 2\phi
  \right)}q^{2}r^{2} + p^{2}t^{2} \right. \nonumber \\ & & \left. -
4e^{i\alpha}pqt^{2} + e^{2i\alpha}q^{2}t^{2} -
3e^{2i\phi}p^{2}\left(r^{2} + t^{2}\right) - 3e^{2i\left(\alpha + \phi
  \right)}q^{2}\left(r^{2} + t^{2}\right)\right],  \nonumber \\ 
y_{4,-} & = & -\frac{1}{2}\;e^{-i\left(2\alpha + 3\phi
  \right)}rt\left[e^{i\left(3\alpha + 4\phi \right)}p^{2}qr^{2} +
  e^{4i\phi}pq^{2}r^{2} - e^{i\left(\alpha + 4\phi
    \right)}q\left(2p^{2} + q^{2} \right)r^{2} \right. \nonumber \\ &
& \left. - e^{2i\left(\alpha + 2\phi \right)}p\left(p^{2} +
    2q^{2}\right)r^{2} +  e^{3i\alpha}p^{2}qt^{2} + pq^{2}t^{2} -
  e^{i\alpha}q\left(2p^{2} + q^{2}\right)t^{2} \right. \nonumber \\ &
& \left. - e^{2i\alpha}p\left(p^{2} + 2q^{2}\right)t^{2} -
  3e^{i\left(3\alpha + 2\phi \right)}p^{2}q\left(r^{2} + t^{2}\right)
  - 3e^{2i\phi}pq^{2}\left(r^{2} + t^{2}\right) \right. \nonumber \\ &
& \left. - e^{i\left(\alpha + 2\phi \right)}q\left(2p^{2} +
    q^{2}\right)\left(r^{2} + t^{2}\right) - e^{2i\left(\alpha + \phi
    \right)}p\left(p^{2} + 2q^{2}\right)\left(r^{2} + t^{2}\right)
\right], \nonumber \\
y_{5,-} & = & \frac{1}{2}\;e^{-i\left(2\alpha + \phi
  \right)}rt\left[-e^{3i\alpha}p^{2}qr^{2} - pq^{2}r^{2} +
  e^{i\alpha}q\left(2p^{2} + q^{2}\right)r^{2} +
  e^{2i\alpha}p\left(p^{2} + 2q^{2}\right)r^{2} \right. \nonumber \\ &
& \left. - e^{i\left(3\alpha +
      4\phi \right)}p^{2}qt^{2} - e^{4i\phi}pq^{2}t^{2} +
  e^{i\left(\alpha + 4\phi \right)}q\left(2p^{2} + q^{2}\right)t^{2} +
  e^{2i\left(\alpha + 2\phi \right)}p\left(p^{2} + 2q^{2}\right)t^{2}
  \right. \nonumber \\ & & \left. + 3e^{i\left(3\alpha + 2\phi
      \right)}p^{2}q\left(r^{2} + t^{2}\right) +
    3e^{2i\phi}pq^{2}\left(r^{2} + t^{2}\right) + e^{i\left(\alpha +
        2\phi\right)}q\left(2p^{2} + q^{2}\right)\left(r^{2} +
      t^{2}\right) \right. \nonumber \\ & & \left. + e^{2i\left(\alpha + \phi
    \right)}p\left(p^{2} + 2q^{2}\right)\left(r^{2} +
    t^{2}\right)\right], \nonumber \\
y_{6,-} & = & -\frac{1}{2}\;e^{-i\phi}\left(p +
  e^{i\alpha}q\right)rt \left[-4e^{i\alpha}pq\left(r^{2} +
    e^{4i\phi}t^{2}\right) - e^{2i\alpha}q^{2}\left(\left(-1 +
      3e^{2i\phi}\right)r^{2} \right. \right. \nonumber \\ & &
\left. \left. - e^{2i\phi}\left(-3 +
      e^{2i\phi}\right)t^{2}\right) + p^{2}\left(\left(1 -
        3e^{2i\phi}\right)r^{2} + e^{2i\phi}\left(-3 +
        e^{2i\phi}\right)t^{2}\right)\right],  \nonumber \\ 
y_{7,-} & = & -\frac{1}{2}\;e^{-i\left(3\alpha + 2\phi
  \right)}\left(e^{i\alpha}p+
  q\right)\left[e^{2i\alpha}p^{2}r^{2}t^{2} + e^{2i\left(\alpha +
      2\phi \right)}p^{2}r^{2}t^{2} - 4e^{i\alpha}pqr^{2}t^{2}
\right. \nonumber \\ & & \left. - 4e^{i\left(\alpha + 4\phi
    \right)}pqr^{2}t^{2} + q^{2}r^{2}t^{2} + e^{4i\phi}q^{2}r^{2}t^{2}
  - e^{2i\left(\alpha + \phi \right)}p^{2}\left(r^{4} + 4r^{2}t^{2} +
    t^{4}\right)\right. \nonumber \\ & & \left. -
  e^{2i\phi}q^{2}\left(r^{4} + 4r^{2}t^{2} + t^{4}\right) \right], \nonumber \\
y_{8,-} & = & \frac{1}{2}\;e^{-i\left(\alpha + 2\phi
  \right)}\left[-p^{2}qr^{2}t^{2} - e^{4i\phi}p^{2}qr^{2}t^{2} -
  e^{3i\alpha}pq^{2}r^{2}t^{2} - e^{i\left(3\alpha + 4\phi
    \right)}pq^{2}r^{2}t^{2} \right. \nonumber \\ & & \left. +
  e^{2i\alpha}q\left(2p^{2} + q^{2}\right)r^{2}t^{2} +
  e^{2i\left(\alpha + 2\phi \right)}q\left(2p^{2} + q^{2}\right)r^{2}t^{2} +
  e^{i\alpha}p\left(p^{2} + 2q^{2}\right)r^{2}t^{2} \right. \nonumber
\\ & & \left. + e^{i\left(\alpha
    + 4\phi \right)}p\left(p^{2} + 2q^{2}\right)r^{2}t^{2} +
e^{2i\phi}p^{2}q\left(r^{4} + 4r^{2}t^{2} + t^{4}\right)
\right. \nonumber \\ & & \left. +
e^{i\left(3\alpha + 2\phi \right)}pq^{2}\left(r^{4} + 4r^{2}t^{2} +
  t^{4}\right) + e^{i\left(\alpha + 2\phi
  \right)}p\left(4q^{2}r^{2}t^{2} + p^{2}\left(r^{4} +
    t^{4}\right)\right) \right. \nonumber \\ & & \left. +
e^{2i\left(\alpha + \phi \right)}q\left(4p^{2}r^{2}t^{2} + q^{2}\left(r^{4} +
    t^{4}\right)\right)\right]
\label{yso62} 
\end{eqnarray}
where we have (\ref{pq}) and (\ref{rt}).

$\bullet$ {\bf $SU(3)$ sector of $CSO(6,2)$ gauging}

With $\xi=0$, the components of $A_2$ tensor can be represented by
eight different fields  $y_{i,0}(i=1,2, \cdots, 8)$
with degeneracies 
3,3,4,12,12,4,6,12 respectively and given by
(\ref{a2su3}) with the replacement  $y_{i,-} \rightarrow y_{i,0}$
where
their explicit expressions are given by
\begin{eqnarray}
y_{1,0} & = & -\frac{1}{4}\;e^{-i\left(\alpha + 4\phi \right)}\left[3p^{2}q +
  3e^{3i\alpha}pq^{2} - e^{2i\alpha}q\left(2p^{2} + q^{2}\right) -
  e^{i\alpha}p\left(p^{2} + 2q^{2}\right)\right] 
 \left(-e^{2i\phi}r^{2}
+ t^{2}\right)^{2},  \nonumber \\ 
y_{2,0} & = & -\frac{1}{4}\;e^{-i\alpha}\left[3p^{2}q + 3e^{3i\alpha}pq^{2} -
  e^{2i\alpha}q\left(2p^{2} + q^{2}\right) - e^{i\alpha}p\left(p^{2} +
  2q^{2}\right)\right]\left(r^{2} -
e^{2i\phi}t^{2}\right)^{2},\nonumber \\  
y_{3,0} & = & -\frac{3}{4}\;e^{-3i\phi}\left(-1 + e^{2i\phi}\right)\left(p -
  e^{i\alpha}q\right)^{2}\left(p +
  e^{i\alpha}q\right)rt\left(e^{2i\phi}r^{2} - t^{2}\right),\nonumber \\ 
y_{4,0} & = & -\frac{1}{4}\;e^{-i\left(2\alpha + 3\phi \right)}\left(-1 +
  e^{2i\phi}\right)\left[3e^{3i\alpha}p^{2}q + 3pq^{2}
  -e^{i\alpha}q\left(2p^{2} + q^{2}\right) \right. \nonumber \\ & &
\left. - e^{2i\alpha}p\left(p^{2}
    + 2q^{2}\right)\right]
rt\left(e^{2i\phi}r^{2} - t^{2}\right),  \nonumber \\
y_{5,0} & = & -\frac{1}{4}\;e^{-i\left(2\alpha + \phi \right)}\left(-1 +
  e^{2i\phi}\right)\left[3e^{3i\alpha}p^{2}q + 3pq^{2}
  -e^{i\alpha}q\left(2p^{2} + q^{2}\right) \right. \nonumber \\ & &
\left. - e^{2i\alpha}p\left(p^{2}
    + 2q^{2}\right)\right]
rt\left(-r^{2} + e^{2i\phi}t^{2}\right), \nonumber \\
y_{6,0} & = & -\frac{3}{4}\;e^{-i\phi}\left(-1 + e^{2i\phi}\right)\left(p -
  e^{i\alpha}q\right)^{2}\left(p + e^{i\alpha}q\right)rt\left(-r^{2} +
e^{2i\phi}t^{2}\right),\nonumber \\ 
y_{7,0} & = & -\frac{1}{4}\;e^{-i\left(3\alpha + 2\phi \right)} \left(
  -e^{i\alpha}p
+ q\right)^{2}\left(e^{i\alpha}p + q\right)\left[3r^{2}t^{2} +
3e^{4i\phi}r^{2}t^{2} - e^{2i\phi}\left(r^{4} + 4r^{2}t^{2}
  +t^{4}\right) \right],\nonumber \\
y_{8,0} & = & \frac{1}{4}\;e^{-i\left(\alpha + 2\phi \right)}\left(p -
  e^{i\alpha}q\right)\left[-3pqr^{2}t^{2} + 3e^{2i\alpha}pqr^{2}t^{2}
  - 3e^{4i\phi}pqr^{2}t^{2} + 3e^{2i\left(\alpha + 2\phi
    \right)}pqr^{2}t^{2} \right. \nonumber \\ & & \left. +
  e^{i\alpha}r^{2}t^{2} + e^{i\left(\alpha + 4\phi \right)}r^{2}t^{2}
  + e^{i\left(\alpha + 2\phi \right)}\left(r^{4} - 4r^{2}t^{2} + t^{4}\right) +
  e^{2i\phi}pq\left(r^{4} + 4r^{2}t^{2} + t^{4}\right)
\right. \nonumber \\ & & \left. -
  e^{2i\left(\alpha + \phi \right)}pq\left(r^{4} + 4r^{2}t^{2} +
    t^{4}\right)\right]
\label{ycso62} 
\end{eqnarray}
with (\ref{pq}) and (\ref{rt}).

$\bullet$ {\bf $SO(5)$ sector of $SO(5,3)$ gauging}

The components of $A_2$ tensor can be represented by
four different fields with degeneracies  $y_{i,-}(i=1,2,3,4)$
16,16,16,8 respectively and
they look similar to the compact 
case(same multiplicities and same number of fields)
and given by
\begin{eqnarray}
A_{2,1}^{\;\;\;\;256} & = & A_{2,1}^{\;\;\;\;278} = A_{2,2}^{\;\;\;\;165} =
A_{2,2}^{\;\;\;\;187} = A_{2,3}^{\;\;\;\;456} = A_{2,3}^{\;\;\;\;478} =
A_{2,4}^{\;\;\;\;365} = A_{2,4}^{\;\;\;\;387} = A_{2,5}^{\;\;\;\;126}
\nonumber \\ & = & A_{2,5}^{\;\;\;\;346} = A_{2,6}^{\;\;\;\;152} =
A_{2,6}^{\;\;\;\;354} = A_{2,7}^{\;\;\;\;128} = A_{2,7}^{\;\;\;\;348}
= A_{2,8}^{\;\;\;\;172} = A_{2,8}^{\;\;\;\;374} \equiv  y_{1,-}
\nonumber \\
A_{2,1}^{\;\;\;\;375} & = & A_{2,1}^{\;\;\;\;368} = A_{2,2}^{\;\;\;\;486} =
A_{2,2}^{\;\;\;\;457} = A_{2,3}^{\;\;\;\;186} = A_{2,3}^{\;\;\;\;157} =
A_{2,4}^{\;\;\;\;275} = A_{2,4}^{\;\;\;\;268} = A_{2,5}^{\;\;\;\;173}
\nonumber \\ & = &
A_{2,5}^{\;\;\;\;247} = A_{2,6}^{\;\;\;\;138} = A_{2,6}^{\;\;\;\;284} =
A_{2,7}^{\;\;\;\;135} = A_{2,7}^{\;\;\;\;254} = A_{2,8}^{\;\;\;\;163} =
A_{2,8}^{\;\;\;\;246} \equiv y_{2,-} \nonumber \\
A_{2,1}^{\;\;\;\;485} & = & A_{2,1}^{\;\;\;\;476} = A_{2,2}^{\;\;\;\;385} =
A_{2,2}^{\;\;\;\;376} = A_{2,3}^{\;\;\;\;267} = A_{2,3}^{\;\;\;\;258} =
A_{2,4}^{\;\;\;\;167} = A_{2,4}^{\;\;\;\;158} = A_{2,5}^{\;\;\;\;184}
\nonumber \\ & = &
A_{2,5}^{\;\;\;\;283} = A_{2,6}^{\;\;\;\;174} = A_{2,6}^{\;\;\;\;273} =
A_{2,7}^{\;\;\;\;146} = A_{2,7}^{\;\;\;\;236} = A_{2,8}^{\;\;\;\;145} =
A_{2,8}^{\;\;\;\;235} \equiv y_{3,-} \nonumber \\
A_{2,1}^{\;\;\;\;234} & = & A_{2,2}^{\;\;\;\;143} = A_{2,3}^{\;\;\;\;124} =
A_{2,4}^{\;\;\;\;132} = A_{2,5}^{\;\;\;\;678} = A_{2,6}^{\;\;\;\;587} =
A_{2,7}^{\;\;\;\;568} = A_{2,8}^{\;\;\;\;576} \equiv y_{4,-} 
\label{a2so5}
\end{eqnarray} 
where
they have explicit simple form  
\begin{eqnarray}
y_{1,-} & = &
\frac{1}{8\sqrt{uvw}}\left(1 + u^{2}v^{2} + u^{2}w^{2} - v^{2}w^{2}\right),
\nonumber \\
y_{2,-} & = &
\frac{1}{8\sqrt{uvw}}\left(1 + u^{2}v^{2} - u^{2}w^{2} + v^{2}w^{2}\right),
\nonumber \\
y_{3,-} & = &
\frac{1}{8\sqrt{uvw}}\left(1 - u^{2}v^{2} + u^{2}w^{2} + v^{2}w^{2}\right),
\nonumber \\
y_{4,-} & = &
\frac{1}{8\sqrt{uvw}}\left(3 + u^{2}v^{2} + u^{2}w^{2} + v^{2}w^{2}\right)
\label{yso53}
\end{eqnarray}
together with (\ref{uvw}).

$\bullet$ {\bf $SO(5)$ sector of $CSO(5,3)$ gauging}

With $\xi=0$, the components of $A_2$ tensor can be represented by
two different fields  $y_{i,0}(i=1,2)$
with degeneracies 
48,8 respectively and given by
\begin{eqnarray}
A_{2,1}^{\;\;\;\;256} & = & A_{2,1}^{\;\;\;\;278} = A_{2,2}^{\;\;\;\;165} =
A_{2,2}^{\;\;\;\;187} = A_{2,3}^{\;\;\;\;456} = A_{2,3}^{\;\;\;\;478} =
A_{2,4}^{\;\;\;\;365} = A_{2,4}^{\;\;\;\;387} = A_{2,5}^{\;\;\;\;126}
\nonumber \\ & = & A_{2,5}^{\;\;\;\;346} = A_{2,6}^{\;\;\;\;152} =
A_{2,6}^{\;\;\;\;354} = A_{2,7}^{\;\;\;\;128} = A_{2,7}^{\;\;\;\;348}
= A_{2,8}^{\;\;\;\;172} = A_{2,8}^{\;\;\;\;374} =
A_{2,1}^{\;\;\;\;375} \nonumber \\  & = & A_{2,1}^{\;\;\;\;368} =
A_{2,2}^{\;\;\;\;486} = A_{2,2}^{\;\;\;\;457} = A_{2,3}^{\;\;\;\;186}
= A_{2,3}^{\;\;\;\;157} = A_{2,4}^{\;\;\;\;275} =
A_{2,4}^{\;\;\;\;268} = A_{2,5}^{\;\;\;\;173} \nonumber \\ & = &
A_{2,5}^{\;\;\;\;247} = A_{2,6}^{\;\;\;\;138} = A_{2,6}^{\;\;\;\;284} =
A_{2,7}^{\;\;\;\;135} = A_{2,7}^{\;\;\;\;254} = A_{2,8}^{\;\;\;\;163} =
A_{2,8}^{\;\;\;\;246} = A_{2,1}^{\;\;\;\;485} \nonumber \\ & = &
A_{2,1}^{\;\;\;\;476} = A_{2,2}^{\;\;\;\;385} = A_{2,2}^{\;\;\;\;376}
= A_{2,3}^{\;\;\;\;267} = A_{2,3}^{\;\;\;\;258} =
A_{2,4}^{\;\;\;\;167} = A_{2,4}^{\;\;\;\;158} = A_{2,5}^{\;\;\;\;184}
\nonumber \\ & = & A_{2,5}^{\;\;\;\;283} = A_{2,6}^{\;\;\;\;174} =
A_{2,6}^{\;\;\;\;273} = A_{2,7}^{\;\;\;\;146} = A_{2,7}^{\;\;\;\;236}
= A_{2,8}^{\;\;\;\;145} = A_{2,8}^{\;\;\;\;235} \equiv y_{1,0} \nonumber \\
A_{2,1}^{\;\;\;\;234} & = & A_{2,2}^{\;\;\;\;143} = A_{2,3}^{\;\;\;\;124} =
A_{2,4}^{\;\;\;\;132} = A_{2,5}^{\;\;\;\;678} = A_{2,6}^{\;\;\;\;587} =
A_{2,7}^{\;\;\;\;568} = A_{2,8}^{\;\;\;\;576} \equiv y_{2,0} 
\label{a2so5other}
\end{eqnarray}
where we have 
\begin{eqnarray}
y_{1,0}  =  \frac{1}{8\sqrt{uvw}}, \qquad
y_{2,0}  =  \frac{3}{8\sqrt{uvw}} 
\label{ycso53} 
\end{eqnarray}
with (\ref{uvw}).

\vspace{2cm}
\centerline{\bf Acknowledgments} 

This research was supported by 
Korea Research Foundation Grant(KRF-2002-015-CS0006). 
CA thanks Max-Planck-Institut f\"ur Gravitationsphysik, 
Albert-Einstein-Institut where part of this work was undertaken and thanks
C.M. Hull and T. Fischbacher for discussions.

\end{document}